\documentclass[prl,twocolumn,aps,superscriptaddress]{revtex4-1}
\usepackage{graphicx}
\usepackage{amsmath}
\usepackage[colorlinks=true, citecolor=blue, urlcolor=blue, linkcolor = blue]{hyperref}
\usepackage{amssymb}
\usepackage{bm}
\usepackage{cancel}
\usepackage[normalem]{ulem} 
\usepackage{soul}
\usepackage[dvipsnames]{xcolor}
\usepackage{float} 
\usepackage[caption=false]{subfig}
\usepackage{mathtools} 
\usepackage{mathrsfs} 
\usepackage{braket}

\begin{document}
	
	\title{Disorder-free localization in an interacting two-dimensional lattice gauge theory}
	
	\author{P. Karpov}
	\email{karpov@pks.mpg.de}
	\affiliation{Max Planck Institute for the Physics of Complex Systems, N{\"o}thnitzer Stra{\ss}e 38, Dresden 01187, Germany}
	\affiliation{National University of Science and Technology ``MISiS'', Moscow, Russia}
	
	\author{R. Verdel}
	\affiliation{Max Planck Institute for the Physics of Complex Systems, N{\"o}thnitzer Stra{\ss}e 38, Dresden 01187, Germany}
	
	\author{Y.-P. Huang}
	\affiliation{The Paul Scherrer Institute, Forschungsstrasse 111, 5232 Villigen, Switzerland}
	
	\author{M. Schmitt}
	\affiliation{Department of Physics, University of California at Berkeley, Berkeley, CA 94720, USA}
	
	\author{M. Heyl}
	\affiliation{Max Planck Institute for the Physics of Complex Systems, N{\"o}thnitzer Stra{\ss}e 38, Dresden 01187, Germany}
	
	\date{March 10, 2020}
	
	\begin{abstract}
		Disorder-free localization has been recently introduced as a mechanism for ergodicity breaking in low-dimensional homogeneous lattice gauge theories caused by local constraints imposed by gauge invariance.  We show that also genuinely interacting systems in two spatial dimensions can become nonergodic as a consequence of this mechanism. Specifically, we prove nonergodic behavior in the quantum link model by obtaining a rigorous bound on the localization-delocalization transition through a classical correlated percolation problem implying a fragmentation of Hilbert space on the nonergodic side of the transition. We study the quantum dynamics in this system by means of an efficient and perturbatively controlled representation of the wavefunction in terms of a variational network of classical spins akin to artificial neural networks. We identify a distinguishing dynamical signature by studying the propagation of line defects, yielding different light cone structures in the localized and ergodic phases, respectively. The methods we introduce in this work can be applied to any lattice gauge theory with finite-dimensional local Hilbert spaces irrespective of spatial dimensionality.
	\end{abstract}
	
	\maketitle

	{\it Introduction.}
	Systems with local constraints play an important role in various physical contexts ranging from strongly correlated electrons \cite{Rokhsar:1988} and frustrated magnets~\cite{Gingras:2014,Stern:2019} to fundamental theories of matter such as quantum electro- and chromodynamics~\cite{Rothe:book}, where constraints take the form of local gauge symmetries.
	The equilibrium properties of such systems have been extensively studied over the last decades, but only recently their nonequilibrium dynamics has moved into focus.
	In particular, local constraints have emerged as a new paradigm for ergodicity breaking, besides the two known archetypical scenarios caused by localization due to strong disorder or integrability.
	Systems with local constraints can exhibit rare nonergodic eigenstates, termed quantum many-body scars~\cite{Turner:2018,Motrunich:2019}, or extremely slow relaxation~\cite{Horrsen:2015,Pancotti:2019,Lan:2018}, whereas dipole conservation can prevent thermalization of large parts of the spectrum in one-dimensional fractonic systems~\cite{Sala:2019,Pai:2019-PRL,Pai:2019-PRX,Khemani:2019}.
	A particularly generic mechanism for nonergodic behavior is hosted in lattice gauge theories (LGTs) where local constraints emerge naturally due to the local gauge symmetry, leading to an extensive number of local conserved quantities.
	Specifically, this can lead to the absence of ergodicity in 1D LGTs with discrete \cite{Smith:2017a,Smith:2017b} and continuous \cite{Brenes:2018} gauge symmetries or for higher-dimensional systems in the low-energy limit~\cite{Haah:2016} or when they are free~\cite{Smith:2018}.
	It has remained, however, a key challenge to rigorously identify non-ergodic behavior in genuinely interacting quantum matter beyond one dimension.
	
	In this work we show that the 2D $U(1)$ quantum link model features both localized and ergodic phases in the absence of disorder.
	Our proof relies on a mapping onto a classical correlated percolation problem providing a rigorous bound on the localization transition of the quantum model.
	We identify a distinguishing quantum-dynamical signature of the two phases by studying the propagation of an initial line defect, which leads to two different light cone structures.
	For the description of the nonequilibrium dynamics of the system we introduce variational classical networks (vCNs), which provide an efficient and perturbatively controlled representation of the quantum many-body wave function in terms of a network of classical spins akin to artificial neural networks (ANNs).
	The introduced methods can be applied to any LGT with finite local Hilbert spaces irrespective of dimensionality.
	
	
	\begin{figure}[!tb]
		\centering
		\includegraphics{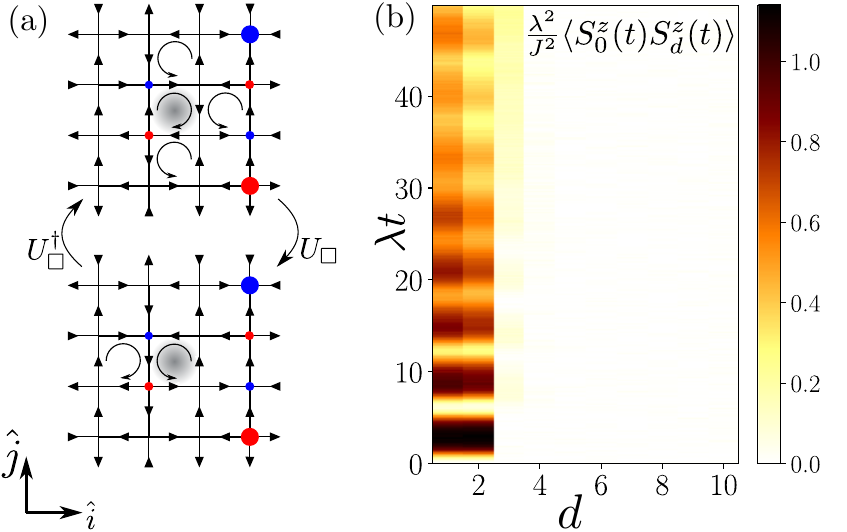}
		\caption{
			a) Illustration of the $U(1)$ quantum link model (QLM) with spin-1/2's located on the links of the square lattice. Spins pointing $\rightarrow$ or $\uparrow$ correspond to $S^z=+1$ and $\leftarrow$ or $\downarrow$ to $S^z=-1$, respectively. Kinetics is introduced by plaquette-flip operators  $U_{\square}, U_{\square}^{\dagger}$ (shown for the darkened central plaquette) whenever the spins on a plaquette are oriented clockwise or counterclockwise. Flippable plaquettes are denoted by circular arrows. Background charges with nonzero in- our outflow of electric field at a given vertex are indicated by red a blue dots. 
			b) Spatiotemporal buildup of quantum correlations $\langle S^z_{\mathbf{r}, y} (t) S^z_{\mathbf{r}+\mathbf{d},y} (t)\rangle \equiv \langle S^z_{0} (t) S^z_{d} (t)\rangle$ ($\mathbf{d} = d\hat{i}$) starting from $|\psi(\alpha=0)\rangle = \ket{\rightarrow}$ in the localized phase of the QLM for $J/\lambda=-0.1$ and a system of size $20\times20$, i.e., $800$ spins.
		}
		\label{fig:model}
	\end{figure}
	
	
	
	{\it Quantum link model.}
	We study the 2D $U(1)$ quantum link model (QLM) \cite{Wiese:1997,Wiese:2013}, which has been introduced as a descendant of lattice quantum electrodynamics with spin-1/2 gauge degrees of freedom.
	In the QLM the spins $S_{\mathbf{r},\mu}$ reside on the links of a square lattice connecting vertices $\mathbf{r} =(x,y)$ and $\mathbf{r}+\mu$  (here $\mu=\hat{i},\hat{j}$ is one of the two unit vectors of the lattice, Fig.~\ref{fig:model}(a)), with the Hamiltonian:
	\begin{align}
	H = H_0 + V \equiv   \lambda \sum_{\square}  (U_{\square} + U_{\square}^{\dagger})^2
	-  J \sum_{\square} (U_{\square} + U_{\square}^{\dagger}).
	\label{eq:HamiltonianQLM}
	\end{align}
	The sums run over all plaquettes $\square$, $U_{\square} = S_{\mathbf{r},\hat{i}}^+ S_{\mathbf{r}+\hat{i},\hat{j}}^+ S_{\mathbf{r}+\hat{j},\hat{i}}^- S_{\mathbf{r},\hat{j}}^-$ induces a collective flip of all spins on plaquette $\square$, and $S^{\pm}_{\mathbf{r},\hat{\mu}}$ denote the raising and lowering operators.
	The first (potential) term counts the number of flippable plaquettes and the second (kinetic) term induces coherent dynamics.
	For what follows, we will consider periodic boundary conditions and the case of a strong potential term with $J/\lambda=-0.1$.
	The QLM not only appears in the context of high-energy physics, but also shares strong connections to condensed matter systems featuring quantum spin ice phases \cite{Shannon:2004,Hermele:2004} or quantum dimer models \cite{Rokhsar:1988,Banerjee:2013}.
	On the experimental side various proposals have explored the potential realization of the QLM in quantum simulators within the last years ~\cite{Celi:2019, Glaetzle:2014}.
	
	 The local gauge symmetry of the QLM is generated by the operators $G_{\mathbf{r}} = \sum_{\mu} (S_{\mathbf{r},\mu}^z - S_{\mathbf{r}-\mu,\mu}^z)$ counting the total inflow of the electric field to the vertex ${\mathbf{r}}$. 
	Since $[G_\mathbf{r}, H] = 0$ for all lattice points and $[G_\mathbf{r},G_{\mathbf{r}'}]=0$, eigenstates of $H$ can be classified by the respective eigenvalues $q_{\mathbf{r}}\in\{-2,-1,0,1,2\}$ of $G_{\mathbf{r}}$. The set of $\mathbf{q}=\{q_{\mathbf{r}}\}$ defines the so-called superselection sector of states $\ket{\psi_{\mathbf{q}}}$ with $G_{\mathbf r} \ket{\psi_{\mathbf{q}}} = q_{\mathbf r} \ket{\psi_{\mathbf{q}}}$, so that each of the $q_\mathbf{r}$ can be given a physical meaning in terms of static background charges located at $\mathbf{r}$~\cite{Brenes:2018}.
	The QLM further has global conserved quantities $\Phi_{x} = \sum_{y} S_{\mathbf{r},\hat{i}}^z$, $\Phi_{y} = \sum_{x} S_{\mathbf{r},\hat{j}}^z$, which define the flux sectors.

	{\it Disorder-free localization.}
	The existence of these sectors, protected by gauge invariance, can lead to an unconventional scenario for ergodicity breaking.
	Consider a homogeneous superposition state $\ket{\psi} = \sum_{\mathbf{q}} C_{\mathbf{q}} \ket{\psi_{\mathbf{q}}}$ involving many superselection sectors.
	As the Hamiltonian and typical observables are block-diagonal, i.e., $H \ket{\psi_{\mathbf{q}}} = H_{\mathbf{q}}\ket{\psi_{\mathbf{q}}}$, the expectation values of an operator $O$ during dynamics become equivalent to $\braket{O(t)} = \sum_{\mathbf{q}} |C_{\mathbf{q}}|^2 \braket{\psi_{\mathbf{q}} | e^{i H_{\mathbf{q}} t}  O e^{-i H_{\mathbf{q}} t} | \psi_{\mathbf{q}}}$ resembling an effective disorder average with the disorder strength determined by the random background charges in the typical superselection sectors~\cite{Brenes:2018}.
	This can, in principle, lead to nonergodic behavior of $\braket{O(t)}$, although both the initial state and the Hamiltonian are homogeneous leading to the notion of disorder-free localization~\cite{Smith:2017a}.
	
	{\it Initial states for time-evolution.}
	We now aim to characterize the nonequilibrium dynamics of the QLM for the following homogeneous initial states $|\psi_0\rangle$:

	(i) $|\psi_0 \rangle = \ket{\rightarrow} = \bigotimes_i \tfrac{1}{\sqrt{2}} (\ket{\uparrow_i}  +   \ket{\downarrow_i})$, where $\ket{\uparrow_i}$ and $\ket{\downarrow_i}$ are the two basis states at link $i$. This state is distributed over all superselection sectors of the model.
	
	(ii) $|\psi_0 \rangle = \ket{\rightarrow}_{\mathrm{FF}}$ which is a projection of $\ket{\rightarrow}$ to a single ``fully-flippable'' (FF) sector, defined as the zero-charge zero-flux sector. $\ket{\rightarrow}_{\mathrm{FF}}$ is an equal-weighted superposition of all states from the FF sector (i.e. the Rokhsar-Kivelson state \cite{Rokhsar:1988} for the FF sector).
	
	While $\ket{\rightarrow}$ is a product state, $\ket{\rightarrow}_{\mathrm{FF}}$ is entangled. Nevertheless, $\ket{\rightarrow}$ can be continuously connected to a product state from the same FF sector via 
	\begin{align}
	|\psi_0 (\alpha) \rangle = \bigotimes_i [\cos(\alpha+\tfrac{\pi}{4}) |\mathrm{FF}_i\rangle + \sin (\alpha+\tfrac{\pi}{4})| \overline{\mathrm{FF}}_i\rangle ] \, ,
	\label{eq:alpha-state}
	\end{align}
	with $\alpha \in [0,\pi/4]$.
	Here, $\ket{\mathrm{FF}_i}$  and $\ket{\mathrm{\overline{FF}}_i}$ denote the local spin orientations of the two states with all plaquettes flippable and therefore with checkerboard-alternating clockwise ($\mathrm{FF}_i$) and anticlockwise ($\mathrm{\overline{FF}}_i$) orientations.
	By construction, $|\psi(\alpha=0)\rangle = \ket{\rightarrow}$ and $|\psi(\alpha=\pi/4)\rangle =\ket{\mathrm{FF}} \equiv \bigotimes_i \ket{\mathrm{FF}_i}$.
	Importantly, the states $|\psi(\alpha)\rangle$ are spatially uniform. 
	The resulting dynamics for $\alpha=0$ is displayed in Fig.~\ref{fig:model}(b), where we monitor the spatiotemporal buildup of quantum correlations.
	We will identify the limited spatial propagation with nonergodic behavior below.
	\\
	
	{\it Variational classical networks.} 
	Calculating the dynamics of interacting quantum systems in 2D is an inherently hard problem without a general-purpose computational method available to date.
	Representing a generic quantum many-body state as $|\psi\rangle = \sum_{\vec{s}} \psi(\vec{s}) \ket{\vec{s}}$ requires, in principle, the storage of exponentially many amplitudes $\psi(\vec{s})$ (here  $\vec{s} = (s_1,...,s_N)$).
	Recently, it has been proposed to use networks of classical spins to solve this problem~\cite{Carleo:2017,Schmitt:2018} by avoiding to store the $\psi(\vec s)$'s.
	The amplitudes $\psi(\vec{s}) \approx  \exp [\mathscr{H} (\vec{s}, \mathcal{W})]$ are rather generated on the fly when needed via a complex classical spin model with Hamiltonian $\mathscr{H} (\vec{s}, \mathcal{W})$ determined by a set of couplings $\mathcal{W}$ between the involved spins.
	Here, we construct $\mathscr{H} (\vec{s}, \mathcal{W})$ using a perturbatively controlled expansion and extend the recently proposed classical networks~\cite{Schmitt:2018} upon imposing an additional optimization principle.
	The resulting approach can be interpreted as encoding $|\psi\rangle$ in an ANN with a specific simplified network structure.

	Within the vCNs we perform an expansion around a classical limit, which in the case of the QLM is the potential term $H_0$ in (\ref{eq:HamiltonianQLM}).
	By representing the evolution operator in the interaction picture $W(t)= \mathcal{T} \exp(-\int_0^t dt' V(t'))$ we can write $\psi(\vec{s},t) = \bra{\vec{s}} e^{iHt}\ket{\psi_0} = \bra{\vec{s}} e^{iH_0 t} W(t)\ket{\psi_0} = e^{iE_{\vec{s}}t} \bra{\vec{s}} W(t)\ket{\psi_0}$, where $H_0 | \vec{s} \rangle =  E_{\vec{s}}| \vec{s} \rangle$.
	For the remaining term $\bra{\vec{s}} W(t)\ket{\psi_0}$ we perform a cumulant expansion for time-ordered exponential operators~\cite{Kubo:1962,Schmitt:2018,SM}, which, e.g., to the first order yields:
 	$\mathscr{H} = -i E_{\vec{s}} t - \bra{\vec{s}} \int_0^t dt' V(t')\ket{\psi_0}/\braket{\vec{s}|\psi_0}$.
 	Taking $|\psi_0\rangle=\ket{\rightarrow}$ one obtains for the QLM
	$\mathscr{H}(\vec{s},t) = -i E_{\vec{s}} t  -i  J\sum^{'}_{\square}  \int_0^t dt'  e^{i  \lambda \omega_{\square}(\vec{s}) t'}$.
	Here $\sum'_{\square}$ denotes the sum over all flippable plaquettes in the spin configuration $\vec s$,
	and $\omega_{\square}=-4,...,4$ counts the difference between number of flippable plaquettes surrounding the given $\square$   before and after its flip (i.e. $\lambda \omega_{\square}$ gives the potential energy difference before and after the flip). For example, for the configuration in Fig.~\ref{fig:model}(a) we have $\omega_{\square} = 3-1=2$  for the central plaquette.
	Going beyond previous work~\cite{Schmitt:2018} we promote $\lambda t$ and its functions such as $\int_0^t dt'  e^{i \lambda \omega_{\square} t'}$ to ten variational parameters $\mathcal{W}_k(t) = (\mathcal{W}^{(0)}, \mathcal{W}^{(1)}_{-4},..., \mathcal{W}^{(1)}_{4})$ yielding $\mathscr{H} (\vec{s}, \mathcal{W}_k(t)) =   -i E_{\vec{s}} \mathcal{W}^{(0)}  -i  J\sum^{'}_{\square}    \mathcal{W}^{(1)}_{\omega_{\square}(\vec{s})}$.
	The local connectivity of the vCN is encoded in the function $\omega_{\square}(\vec{s})$.
	For the actual shown numerical simulations we use a second-order ansatz and more complex initial states (see \cite{SM,Verdel:2020}).
	The $\mathcal{W}_k(t)$'s are determined by a time-dependent variational principle translating quantum dynamics into a system of coupled classical differential equations $\sum_{k'}\mathcal{S}_{k,k'} \dot{\mathcal{W}}_{k'} = -i F_{k}$ in the space of variational parameters.
	Here, $\mathcal{S}_{k, k'}=\langle O_k^{\ast} O_{k'}\rangle-\langle O_k^{\ast} \rangle\langle O_{k'}\rangle$ and $F_k=\langle E_\mathrm{loc}O_k^{\ast}\rangle - \langle E_\mathrm{loc}\rangle \langle O_k^{\ast}\rangle$, 
	with $O_k(\vec{s})= \partial \ln \psi(\vec{s}, \mathcal{W}) / \partial \mathcal{W}_k$ and 
		 $E_\mathrm{loc}(\vec{s})=\langle \vec{s}|H|\psi_{\mathcal{W}}\rangle  / \langle\vec{s}|\psi_{\mathcal{W}} \rangle$ (see \cite{SM}).
	We solve these equations using a 4th-order Runge-Kutta integrator with step size $\Delta t=0.1 \lambda^{-1}$ and sample the observables using Metropolis Monte Carlo (MC) with $10^6$ sweeps at each time instance, with single spin-flip updates for $\ket{\rightarrow}$ and plaquette flips for $\ket{\rightarrow}_{\mathrm{FF}}$.
	
	While our approach is numerically stable and therefore doesn't face some challenges appearing in ANNs~\cite{Schmitt:2019}, it shares its own limitations due to its perturbative construction,
	which is guaranteed to work only up to times $t \simeq |1/J|$.
	We find, however, that the errors remain perturbatively controlled up to much longer times as a consequence of the variational optimization.
	This can be verified, since the method provides a self-contained way of tracking the error, not referring to any reference solution. We present the details of the error analysis in \cite{SM,Verdel:2020} together with benchmarks.


	\begin{figure}[!t]
		\centering
		\includegraphics{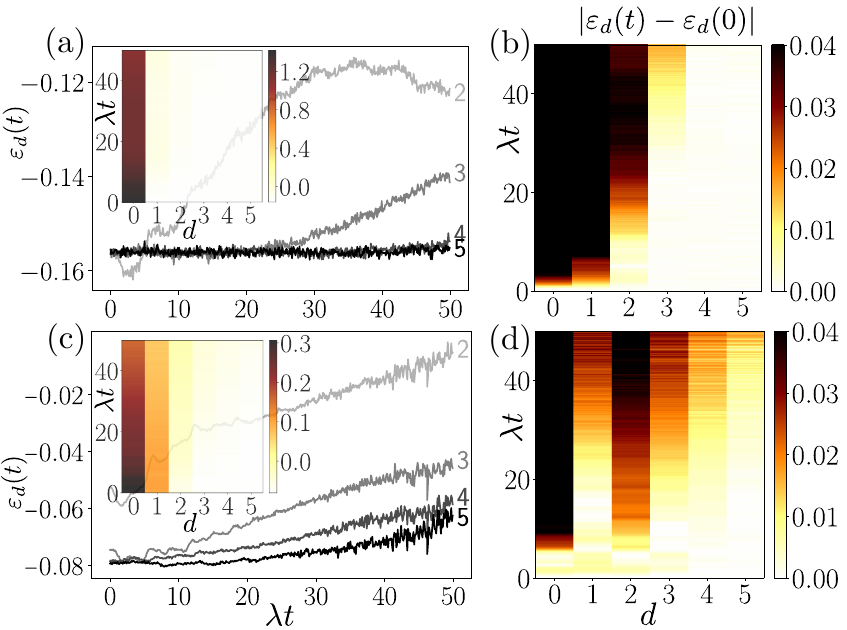}
		\caption{
			Quantum energy dynamics for line defects created in the $\ket{\rightarrow}$ (upper panel) and $\ket{\rightarrow}_{\mathrm{FF}}$ (lower panel) initial states, displaying the normalized plaquette energy $\varepsilon_d(t)$ of $d$-th column.
			(a,c) $\varepsilon_d(t)$  for different columns $d=2\div5$ where darker colors refer to larger distances from the initial defect. Insets show the same data including $d=0,1$ in a color plot.
			(b,d) Absolute deviation of $\varepsilon_d(t)$ from the initial value $\varepsilon_d(0)$.
			(b) Signal propagation for $\ket{\rightarrow}$ showing a strong bending of the light cone indicating localized behavior and (d) for $\ket{\rightarrow}_{\mathrm{FF}}$ consistent with linear propagation indicative of ergodic behavior. For all the plots $J/\lambda =-0.1$ and  system size $10\times10$.
		}
		\label{fig:quantum-dynamics}
	\end{figure}
	
	
	\begin{figure*}[!tbh]
		\centering
		\includegraphics{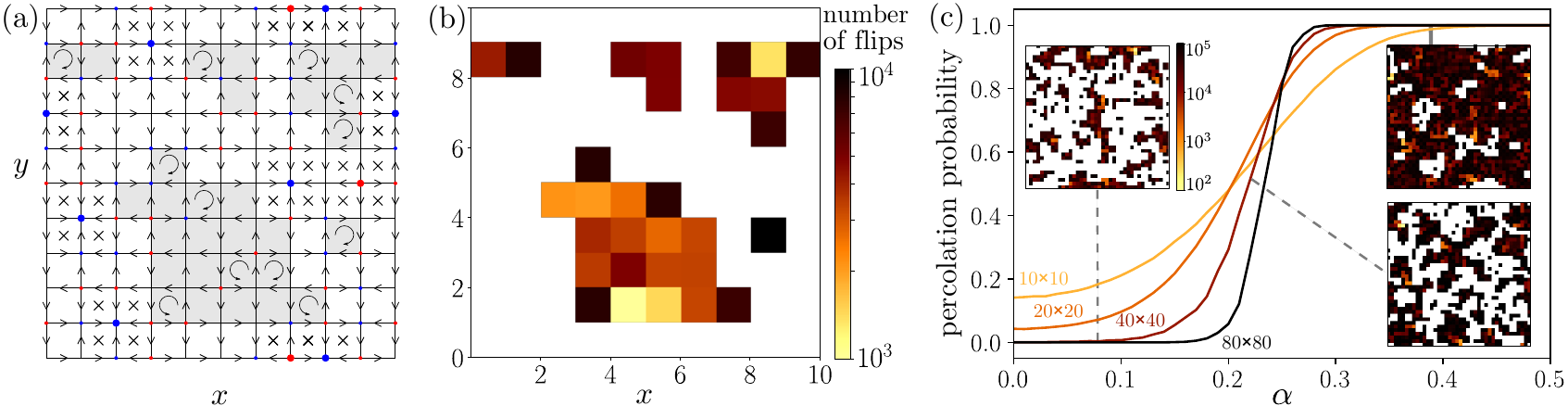}
		\caption{
			Classical correlated percolation problem implying a quantum nonergodic phase in the QLM.
			(a) Typical spin configuration sampled from $|\rightarrow\rangle$ with arrows showing the local spin orientation and circular arrows flippable plaquettes. Red and blue disks denote positive and negative background charges, respectively. Crosses indicate plaquettes blocked by charges $q=\pm2$. Grey color indicates plaquettes that can become flippable in the course of the evolution. 
			(b)  Result of the Monte Carlo simulation starting from the state (a). The colormap shows the number of times that the individual plaquettes were flipped in the course of the simulation; white color stands only for plaquettes, that have never been flipped.
			c) Percolation probability vs $\alpha$.
			The insets show typical configurations below, at, and above  the percolation threshold $\alpha_c\approx 0.25$ for a $40\times40$ system.
		}
		\label{fig:percolation-threshold}
	\end{figure*}

	
	{\it Localized  and ergodic dynamics.} 
	Using the vCNs we now compute nonequilibrium dynamics in the QLM.
	We start by studying the spatiotemporal buildup of quantum correlations, measured via $\langle S^z_0 (t) S^z_d (t)\rangle$, upon initializing the system in state $\ket{\psi_0}  = \ket{\rightarrow}$.
	The result is shown in Fig.~\ref{fig:model}(b), where one can see that correlations emerge only over a limited spatial distance suggesting nonergodic behavior.
	We proceed by further corroborating this observation by other measures.
	
	Namely, we study energy transport in the QLM by creating initial conditions with a spatial energy inhomogeneity in the form of a line defect with subextensive energy contribution and use the character of energy propagation to distinguish between ergodic and localized dynamics.
	%
	%
	Concretely, we consider the two initial conditions $\ket{\psi_0} = \ket{\rightarrow}$ or $\ket{\rightarrow}_{\mathrm{FF}}$ upon applying in addition $P = \prod_{\square \in \mathcal{C}_0}  [1+ (U_{\square} + U_{\square}^{\dagger})^2]$ along all plaquettes in column $d=0$; here $\mathcal{C}_d$ denotes the set of plaquettes in column $d$.
	In Fig.~\ref{fig:quantum-dynamics} we plot the (normalized) column energy $\varepsilon_d(t) = (\langle H_d(t)\rangle - H_{\mathrm{av}})/H_{\mathrm{av}}$ with $H_d = \sum_{\square\in \mathcal{C}_d} H_\square$ the total energy for the plaquettes in $d$-th column, $H_{\square} = \lambda  (U_{\square} + U_{\square}^{\dagger})^2 - J (U_{\square} + U_{\square}^{\dagger})$.
	Further, $H_{\mathrm{av}} = \langle H \rangle / L$ denotes the expected $\langle H_d(t)\rangle$ in the long-time limit when the system is thermalizing ($L$ is the number of columns).

	Comparing Figs.~\ref{fig:quantum-dynamics}(a) and (c) we observe that the dynamics differs qualitatively for the two initial conditions, although the Hamiltonian parameters are identical.
	While for $|\psi_0\rangle = \ket{\rightarrow}$ energy transport is highly suppressed and only visible on short distances (Fig.~\ref{fig:quantum-dynamics}(a)), the opposite happens for $\ket{\rightarrow}_{\mathrm{FF}}$.
	This becomes even more apparent in Figs.~\ref{fig:quantum-dynamics}(b,d), where $\varepsilon_d(t)$ relative to the initial value $\varepsilon_d(0)$ is shown, therefore more directly highlighting energy propagation.
	While for $\ket{\rightarrow}_{\mathrm{FF}}$ we identify a linearly propagating front, for $\ket{\rightarrow}$ we observe a strong bending.
	We argue below that this front for $\ket{\rightarrow}$ can extend only to a finite region as a consequence of disorder-free localization.
	\\
	
	
	{\it Bound on quantum dynamics by unconventional percolation.} 
	The qualitative difference in the quantum dynamics for the initial states $\ket{\rightarrow}$ and $\ket{\rightarrow}_{\mathrm{FF}}$ originates from a dynamical transition, which one can study systematically upon tuning the parameter $\alpha$ for the initial state (\ref{eq:alpha-state}).
	For this purpose, we employ an unconventional correlated classical percolation problem \cite{Coniglio:2009}  and establish a bound on the quantum localized-ergodic transition in the QLM providing a numerical proof for an extended nonergodic phase as a consequence of disorder-free localization.
	
	We illustrate the idea for the initial state $\ket{\rightarrow}$, distributed over all superselection sectors.
	Consider a typical (random) sector from this distribution (Fig.~\ref{fig:percolation-threshold}(a)).
	Such sector exhibits many background charges $q_{\mathbf{r}}$ whenever the ``two-in two-out'' rule at vertex $\mathbf{r}$ is violated.
	Importantly, these background charges (constants of motion by gauge invariance) impose strong kinetic constraints.
	For instance, $q_{\mathbf{r}}=\pm2$ implies that neighboring spins either all point inwards or outwards, hence
	the adjacent plaquettes remain unflippable forever.
	The influence of $q_{\mathbf{r}}=\pm1$ charges is more subtle.
	They make at least $2$ adjacent plaquettes unflippable, while their positions might change over time.

	The question we address now is whether these constraints are so strong to fragment the square lattice into sets of kinetically disconnected islands or whether one can contain an extensive (percolating) connected cluster. 
	For that purpose we study an unconventional percolation problem using an infinite-temperature classical MC simulation.
	We start from the initial condition (\ref{eq:alpha-state}), sampling a random basis state (and thus a sector) with a distribution set by the amplitudes in $|\psi(\alpha)\rangle$.
	Then we determine which parts of the systems are kinetically connected, using MC search with random plaquette flips.
	The simulation is stopped when every plaquette is flipped either 0 or more than some fixed threshold ($=100$) number of times (or after $10^{11}$ MC steps if this condition is still not satisfied).
	As a result we find the number of performed flips for each plaquette (Fig.~\ref{fig:percolation-threshold}(b)). 
	Repeating this procedure for different initial configurations at a given $\alpha$ and scanning $\alpha$, we finally obtain the percolation probability (Fig.~\ref{fig:percolation-threshold}(c)).
	Most importantly, one can observe a clear evidence for a percolation threshold $\alpha_c \approx 0.25$.
	Although the simulation termination condition is chosen such as to minimize the number of potentially missed ``weak connections'' between flippable clusters, we cannot exclude the possibility of such misses.
	While we don't expect a significant impact deep in the respective phases, this caveat might become important in the vicinity of $\alpha_c$, preventing us to obtain a precise value of $\alpha_c$ and to study the critical behavior.
	
	Since $\alpha_c>0$, the initial state $\ket{\psi(\alpha=0)} = \ket{\rightarrow}$ corresponds to the classically non-percolating side of the transition, while from $\alpha_c < \pi/4$ it follows that state $\ket{\psi(\alpha=\pi/4)} =\ket{\mathrm{FF}}$ and all other states from the FF-sector (including $\ket{\rightarrow}_{\mathrm{FF}}$) 	lie on the percolating side. 
	This classical threshold is imprinted in the quantum dynamics and ultimately leads to the strong localization observed in propagation of correlations (Fig.~\ref{fig:model}(b)) and of the energy (Fig.~\ref{fig:quantum-dynamics}(a),(b)) for $\ket{\rightarrow}$. For the FF-sector state $\ket{\rightarrow}_{\mathrm{FF}}$ there is no percolation constraint, which allows propagation of the signal to long distances (Fig.~\ref{fig:quantum-dynamics}(c),(d)).
	This analysis sets a lower bound onto the critical value of the quantum transition $\alpha_c^{(q)}$, since the quantum system might be still localized due to interference even on the classically percolating side.

	
	{\it Summary and outlook.} 
	We have shown that genuinely interacting 2D homogeneous LGTs can become nonergodic as a consequence of disorder-free localization.
	This is all the more surprising as many-body localization is predicted to be unstable in 2D at elevated energy densities~\cite{Deroeck:2017}, implying that gauge invariance represents a more robust mechanism of ergodicity breaking compared to conventional disorder.
	The key element of our analysis is a bound on the localization-delocalization transition based on a classical correlated percolation problem implying a strong fragmentation of Hilbert space into kinetically disconnected regions.
	Both the percolation analysis as well as the  introduced variational classical networks can be directly applied to other quantum many-body systems with finite-dimensional local Hilbert spaces independent of dimensionality, such as 3D quantum spin ice systems, which might be an interesting scope of the developed techniques in the future.
	Further, it might be interesting to explore how the quantum and classical percolation thresholds are related to each other as well as to determine their respective critical behaviors, and  whether the disorder-free localization scenario holds also in the presence of matter degrees of freedom.
	Our theoretical analysis appears within reach of future experiments: significant efforts in the last years have explored routes to realize the QLM model experimentally in systems of Rydberg atoms~\cite{Celi:2019, Glaetzle:2014} as a next step after the recent experimental advances on 1D LGTs~\cite{Martinez:2016,Kokail:2019,Esslinger:2019,Aidelsburger:2019,Mil:2019}.
	%

	
	{\it Acknowledgments.} We are grateful to H. Burau, J. Chalker, M. Dalmonte, R. Moessner, and G.-Y. Zhu for helpful discussions. P.K. acknowledges the support of the Alexander von Humboldt Foundation and the Ministry of Science and Higher Education of the Russian Federation.	This project has received funding from the European Research Council (ERC) under the European Union’s Horizon 2020 research and innovation programme (grant agreement No. 853443), and M. H. further acknowledges support by the Deutsche Forschungsgemeinschaft via the Gottfried Wilhelm Leibniz Prize program. Y.-P.H. receives funding from the European Union's Horizon 2020 research and innovation program under the Marie Sk\l odowska-Curie grant agreement No. 701647. M.S. was supported through the Leopoldina Fellowship Programme of the German National Academy of Sciences Leopoldina (LPDS 2018-07) with additional support from the Simons Foundation.



\begin{thebibliography}{99}
		

		\bibitem{Rokhsar:1988} D.S. Rokhsar  and S.A. Kivelson,  \textit{Phys. Rev. Lett.} \textbf{61}, 2376 (1988).
		
		\bibitem{Gingras:2014} M.J.P. Gingras and P.A. McClarty,  \textit{Rep. Prog. Phys.} 77, 056501 (2014)
		
		\bibitem{Stern:2019} M. Stern, C. Castelnovo, R. Moessner, V. Oganesyan, and S. Gopalakrishnan, arXiv 1911.05742.
		
		 \bibitem{Rothe:book} Lattice Gauge Theories: An Introduction (3nd ed.), H.J.~Rothe (World Scientific, 2005).
		
		\bibitem{Turner:2018} C.J. Turner, A.A. Michailidis, D.A. Abanin, M. Serbyn, and Z. Papic,  \textit{Nat. Phys.} \textbf{14}, 745 (2018)
		\bibitem{Motrunich:2019} C.-J. Lin and O.I. Motrunich, \textit{Phys. Rev. Lett.} \textbf{122}, 173401 (2019)
		
		\bibitem{Horrsen:2015} M. van Horssen, E. Levi, and J.P. Garrahan, \textit{Phys. Rev. B} \textbf{92}, 100305(R) (2015).
		\bibitem{Pancotti:2019} N. Pancotti, G. Giudice, J I. Cirac, J.P. Garrahan, and M.C. Banuls, arXiv:1910.06616.
		
		\bibitem{Lan:2018} Z. Lan, M. van Horssen, S. Powell, and J.P. Garrahan,  \textit{Phys. Rev. Lett.} \textbf{121},  040603  (2018).
		
		\bibitem{Sala:2019} P. Sala, T. Rakovszky, R. Verresen, M. Knap, and F. Pollmann,  arXiv 1904.04266.
		
		\bibitem{Pai:2019-PRL} S. Pai and M. Pretko,  \textit{Phys. Rev. Lett.} \textbf{123}, 136401 (2019).
		\bibitem{Pai:2019-PRX} S. Pai, M. Pretko, and R.M. Nandkishore,  \textit{Phys. Rev. X} \textbf{9},  021003 (2019).
		\bibitem{Khemani:2019} V. Khemani and R. Nandkishore, arXiv:1904.04815.
		
		
		\bibitem{Smith:2017a} A. Smith, J. Knolle, D.L. Kovrizhin, and R. Moessner, \textit{Phys. Rev. Lett.} \textbf{118},  266601 (2017).
		
		\bibitem{Smith:2017b} A. Smith, J. Knolle, R. Moessner, and D.L. Kovrizhin, \textit{Phys. Rev. Lett.} \textbf{119},  176601 (2017).
		
		\bibitem{Brenes:2018} M. Brenes, M. Dalmonte, M. Heyl, and A. Scardicchio, \textit{Phys. Rev. Lett.} \textbf{120},   030601 (2018).
		
		
		\bibitem{Haah:2016} I.H. Kim and J. Haah, \textit{Phys. Rev. Lett.} \textbf{116}, 027202 (2016).
		
		\bibitem{Smith:2018} A. Smith, J. Knolle, R. Moessner, and D.L. Kovrizhin, \textit{Phys. Rev. B} \textbf{97},  245137 (2018).
		
		
		\bibitem{Wiese:2013} U.-J. Wiese,  \textit{Annalen der Physik} \textbf{525}, 777 (2013).
		
		\bibitem{Wiese:1997} S. Chandrasekharan and U.-J. Wiese,  \textit{Nuclear Physics B} \textbf{492}, 455 (1997).
		
		\bibitem{Shannon:2004} N. Shannon, G. Misguich, and K. Penc, \textit{Phys. Rev. B} \textbf{69},  220403(R) (2004).
		
		\bibitem{Hermele:2004} M. Hermele, M.P.A. Fisher, and L. Balents, \textit{Phys. Rev. B} \textbf{69},  064404 (2004). 
		
		\bibitem{Banerjee:2013} D. Banerjee, F.-J. Jiang, P. Widmer, and U.-J. Wiese, \textit{J.~Stat. Mech} P12010 (2013).
		
		\bibitem{Celi:2019} A. Celi, B. Vermersch, O. Viyuela, H. Pichler, M.D. Lukin, P. Zoller, arXiv 1907.03311.
		\bibitem{Glaetzle:2014} A. Glaetzle, M. Dalmonte, R. Nath, I. Rousochatzakis, R. Moessner, and P. Zoller, \textit{Phys. Rev. X} \textbf{4}, 041037 (2014).
		
		\bibitem{Carleo:2017} G. Carleo, M. Troyer,  \textit{Science} \textbf{355}, 602 (2017).

		\bibitem{Schmitt:2018}  M. Schmitt and M. Heyl, \textit{SciPost Phys.} \textbf{4}, 013 (2018).

		\bibitem{Kubo:1962} R. Kubo, Generalized Cumulant Expansion Method, \textit{J. Phys. Soc. Jpn.} \textbf{17}, 1100 (1962).
		
		\bibitem{SM} See the Supplementary Material.
		
		\bibitem{Verdel:2020} R. Verdel, P. Karpov, Y.-P. Huang, M. Schmitt and M. Heyl, to appear. 
		
		\bibitem{Schmitt:2019}  M. Schmitt and M. Heyl, arXiv:1912.08828.

		\bibitem{Coniglio:2009} A. Coniglio, A. Fierro.  Correlated Percolation. In: Meyers R. (eds) Encyclopedia of Complexity and Systems Science (Springer, New York, 2009).

		\bibitem{Deroeck:2017} W. De Roeck and J. Z. Imbrie, \textit{Philos. Trans. Royal Soc. A} \textbf{375}, 20160422 (2017).
		
		\bibitem{Martinez:2016}  E. Martinez, C. Muschik, P. Schindler, D. Nigg, A. Erhard, M. Heyl, P. Hauke, M. Dalmonte, T. Monz,
		P. Zoller, and R. Blatt, \textit{Nature} \textbf{534}, 516 (2016).
		\bibitem{Kokail:2019} C. Kokail, C. Maier, R. van Bijnen, T. Brydges, M. K. Joshi, P. Jurcevic, C. A. Muschik, P. Silvi, R. Blatt, C. F. Roos, et al., Nature  569  355, (2019). 
		\bibitem{Esslinger:2019} F. G{\"o}rg, K. Sandholzer, J. Minguzzi, R. Desbuquois , M. Messer and
		T. Esslinger, \textit{Nature Physics}  \textbf{15}, 1161 (2019).
		\bibitem{Aidelsburger:2019} C. Schweizer, F. Grusdt, M. Berngruber, L. Barbiero, E. Demler, N. Goldman, I. Bloch, and M. Aidelsburger, \textit{Nature Physics}  \textbf{15}, 1168 (2019).
		\bibitem{Mil:2019} A. Mil, T.V. Zache, A. Hegde, A. Xia, R.P. Bhatt, M.K. Oberthaler, P. Hauke, J. Berges, and F. Jendrzejewski, arXiv:1909.07641.
		
		
	\end{thebibliography}
\end{document}


\title{Supplemental Material for ``Disorder-free localization in an interacting 2D lattice gauge theory''}
	
	\author{P. Karpov}
	
	\author{R. Verdel}
	
	\author{Y.-P. Huang}
	
	\author{M. Schmitt}
	
	\author{M. Heyl}
	
	\date{March 10, 2020}
	\maketitle
	\onecolumngrid
	%
	%
	\vspace{-1cm}
	\section{Construction of variational classical network \it{Ans{\"a}tze}}
	\label{Construction of variational classical network Ansatze}
	
	In this Supplementary Section we outline the procedure of construction of a variational classical networks for the quantum link model, which can be straightforwardly generalized to other quantum models with finite local Hilbert spaces [\onlinecite{Verdel:2020}].
	
	If the Hamiltonian of the model can be written as $H = H_0 + V$, where $H_0$ is diagonal in the computational basis and $V \ll H_0$ is a small non-diagonal perturbation, then in order to describe the evolution of the wave function we can use a perturbative treatment based on the cumulant expansion [\onlinecite{Schmitt:2018}].
	Moreover, we can build on top of this perturbative description and construct adequate variational \textit{Ans{\"a}tze} for the wavefunction.
	
	For the studied quantum link model we have $H = H_0 + V \equiv \lambda \sum_{\square}  (U_{\square} + U_{\square}^{\dagger})^2
	-  J \sum_{\square} (U_{\square} + U_{\square}^{\dagger})$, where the sums run over all plaquettes $\square$.
	In the perturbative limit it is convenient to work in the interaction representation and rewrite the evolution operator as $e^{-iHt} = e^{-iH_0 t} W(t)$, where $W(t) = \mathcal{T} \exp (-i \int_0^t dt' V(t'))$.
	Let further  the initial state be $|\psi_0\rangle = \sum_{\vec{s}} \psi_0(\vec{s}) \ket{\vec{s}}$ (i.e. $\psi_0(\vec{s}) = \langle \vec{s}| \psi_0\rangle$), where $|\vec{s}\rangle = |s_1, s_2, ...\rangle$; and the time-evolved state $|\psi (t)\rangle = e^{iHt} \ket{\psi_0} = \sum_{\vec{s}} \psi(\vec{s},t) | \vec{s} \rangle$.
	
	Below we explicitly describe the construction of the first-order ansatz and sketch an analogous derivation for the second-order ansatz. 
	
	\subsection{First-order variational ansatz}

	Within the above settings the coefficients for the time-evolved state $\ket{\psi(t)}$ in the lowest-order in the cumulant expansion [\onlinecite{Kubo:1962}] in the perturbation $V$ can be obtained as
	\begin{align}
	\psi(\vec{s},t) = \bra{\vec{s}} e^{iH_0 t} W(t) \ket{\psi_0} =  e^{i E_{\vec{s}} t} \bra{\vec{s}} W(t) \ket{\psi_0} = e^{i E_{\vec{s}} t} \bra{\vec{s}} 1- i  \int_0^t dt' V(t') + O(V^2))\ket{\psi_0} = \nonumber \\
	= \braket{\vec{s}|\psi_0} e^{i E_{\vec{s}} t} \left(1 - i \frac{\bra{\vec{s}}  \int_0^t dt' V(t') \ket{\psi_0}}{\braket{\vec{s}|\psi_0}} + O(V^2)\right),
	\label{eq:Cumulant_expansion-0}
	\end{align}
	where $E_{\vec{s}}$ is the eigenvalue of the unperturbed Hamiltonian: $H_0 | \vec{s} \rangle =  E_{\vec{s}}| \vec{s} \rangle$ and we assumed that $\braket{\vec{s}|\psi_0}$ doesn't contain any special smallness. To the lowest order in $V$ we can reexponentiate the result and obtain
	\begin{align}
	\psi(\vec{s},t) = \psi_0(\vec{s})  e^{-i E_{\vec{s}} t} \exp \left(  -i \int_0^t dt' \frac{\langle \vec{s} | V(t')|\psi_0 \rangle}{\langle \vec{s} | \psi_0 \rangle}  + O(V^2)\right)
	\label{eq:Cumulant_expansion}
	\end{align}
	We recast our wavefunction coefficient in the form of an effective Hamiltonian of a classical spin model $\mathscr{H}$: $\psi(\vec{s},t) = e^{\mathscr{H} (\vec{s},t)}$ convenient for Monte Carlo sampling:
	\begin{align}
	\psi(\vec{s},t) = \exp \left( \ln \psi_0(\vec{s})  -i E_{\vec{s}} t  -i  \sum_{\vec{s}_1} \frac{\psi_0(\vec{s}_1)}{\psi_0(\vec{s})}\int_0^t dt' \langle \vec{s} | V(t')| \vec{s}_1 \rangle   + O(V^2)\right).
	\label{eq:Cumulant_expansion-2}
	\end{align}
	In order to proceed we need to find $V(t)$, which can be expressed using $U_{\square}(t)$.
	Equation of motion for operators $U_{\square}(t) \equiv e^{i H_0 t} U_{\square}e^{-i H_0 t}$ in the interaction representation is given by
	\begin{align}
	-i \frac{d}{dt} U_{\square} (t)  = [H_0, U_{\square}(t)]  
	\label{eq:U_EOM}
	\end{align}
	\begin{figure}[hbt] %
		\centering
		\includegraphics[width=4cm]{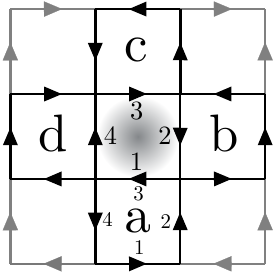}	
		\caption{Convention for numbering the spins within the plaquette (1, 2, 3, 4 are the bottom, right, upper, and left spins) and for labeling the neighboring plaquettes (a, b, c, d) for a given one (central, darkened). Recall that spin directions $\rightarrow$, $\uparrow$ correspond to $S^z=+1$, while $\leftarrow$ and $\downarrow$ correspond to $S^z=-1$ and note that $S_1 \equiv S_{a3}$, $S_2 \equiv S_{b4}$ and so on.}
		\label{fig:Plaquette_notation}
	\end{figure}
	Only four plaquettes (indexed by letters a, b, c, d, Fig. \ref{fig:Plaquette_notation}) adjacent to the plaquette under consideration  $\square$ contribute to the commutator
	$[H_0, U_{\square}] = \lambda \Omega_{\square} U_{\square}$, where $\Omega_{\square}$ is a diagonal operator 
	\begin{align}
	\Omega_{\square}  = 
	-P_{a1}^+ P_{a2}^+ P_{a4}^-  -  P_{b1}^+ P_{b2}^+ P_{b3}^- - P_{c2}^+ P_{c3}^- P_{c4}^-  -  P_{d1}^+ P_{d3}^- P_{d4}^-  \nonumber \\
	+ P_{a1}^- P_{a2}^- P_{a4}^+ + P_{b1}^- P_{b2}^- P_{b3}^+ + P_{c2}^- P_{c3}^+ P_{c4}^+ + P_{d1}^- P_{d3}^+ P_{d4}^+ 
	\end{align}
	and $P^+$ and $P^-$ are projectors to $S^z=+1$ and $S^z=-1$ correspondingly: $P^{\pm} =(1 \pm S^z)/2$.
	Thus the equation of motion (\ref{eq:U_EOM}) transforms to $-i \frac{d}{dt} U_{\square} (t)  = \lambda \Omega_{\square} U_{\square}(t)$ and can be readily solved by
	\begin{align}
	U_{\square}(t) &= e^{i \lambda\Omega_{\square}t} U_{\square} \label{eq:U_t1}\\
	U_{\square}^{\dagger}(t) &= e^{-i \lambda\Omega_{\square}t} U_{\square}^{\dagger}
	\label{eq:U_t2}
	\end{align}
	(note that $\Omega_{\square}$ commutes with $U_{\square}$ since they do not contain the same spins).
	
	Now we're ready to substitute $V(t) = J \sum_{\square} (U_{\square}(t) + U_{\square}^{\dagger}(t))$ in (\ref{eq:Cumulant_expansion-2}). In order to do so it's convenient to introduce the diagonal operator $\mathcal{F}_{\square} = U_{\square}  U_{\square}^{\dagger} -  U_{\square}^{\dagger} U_{\square}$; so its matrix element $\mathcal{F}_{\square}(\vec{s})=+1,-1,0$ is the flippability of the plaquette $\square$ in the state $\vec{s}$ for counterclockwise ($+1$) or clockwise-oriented ($-1$) flippable, or non-flippable plaquettes ($0$) respectively. By $\vec{s}_{\square}$ we denote configuration which differs from $\vec{s}$ by the flip of single plaquette: $|\vec{s}_{\square} \rangle \equiv (U_{\square}+U_{\square}^{\dagger}) | \vec{s} \rangle$. It's also convenient to denote $\omega_{\square}(\vec{s}) =  \mathcal{F}_{\square}(\vec{s})  \Omega_{\square}(\vec{s})$. 
	It takes values $\omega_{\square}=-4,...,4$ and counts the difference between number of flippable plaquettes surrounding the given $\square$   before and after its flip (i.e. $\lambda \omega_{\square}$ gives the potential energy difference before and after the flip).
	Then $\langle \vec{s} | ( e^{i \lambda\Omega_{\square}t} U_{\square} + e^{-i \lambda\Omega_{\square}t} U_{\square}^{\dagger}) | \vec{s}_{\square} \rangle = e^{i \mathcal{F}_{\square}(s) \Omega_{\square}(\vec{s}) t} = e^{i \omega_{\square}(\vec{s}) t}$. Finally substituting $V(t)$ to (\ref{eq:Cumulant_expansion-2}) we get
	\begin{align}
	\psi(\vec{s},t) = \exp \left( \ln \psi_0(\vec{s})  -i E_{\vec{s}} t  -i  J \sideset{}{'}\sum_{\square}   \frac{\psi_0(\vec{s}_{\square})}{\psi_0(\vec{s})}\int_0^t dt'  e^{i \lambda \omega_{\square}(\vec{s})t'}   + O(V^2)\right)
	\label{eq:Cumulant_expansion-3}
	\end{align}
	where prime over the sum indicates summation over only flippable plaquettes. This gives us the lowest-order term of the perturbative expansion of $\psi(\vec{s},t)$.
	
	Now we want to construct a variational ansatz that should at least reproduce this perturbative result (\ref{eq:Cumulant_expansion-3}) and hopefully improve it. 
	We use the following strategy: we promote all functions of time to the variational parameters $\mathcal{W}_k$ in such a way that the effective classical Hamiltonian $\mathscr{H}(\vec{s},\mathcal{W})$ was a linear function of them.
	The non-trivial time dependence sits in the integrals
	\begin{align}
	f(\omega_{\square},t) = \int_0^t dt'  e^{i \lambda \omega_{\square} t'}
	\label{eq:time-VP}
	\end{align}
	Function $f(\omega_{\square},t)$ depends on $t$ and on the discrete parameter $\omega_{\square}$ that takes values $-4,..,4$. We promote this function to the most general function of the same form
	\begin{align}
	f(\omega_{\square}(\vec{s}),t) \equiv \sum_{\omega_1=-4}^4 \delta_{\omega_{\square}(\vec{s}),\omega_1} f(\omega_1, t) \rightarrow 
	\sum_{\omega_1=-4}^4 \delta_{\omega_{\square}(\vec{s}),\omega_1} \mathcal{W} _{\omega_1}^{(1)} = 
	\mathcal{W} _{\omega_{\square}(\vec{s})}^{(1)} 
	\label{eq:time-to-VP}
	\end{align}
	introducing nine first-order variational parameters $\mathcal{W} _{-4}^{(1)}, ..., \mathcal{W} _{4}^{(1)}$. In the initial state all variational parameters are zero.
	Finally along with the 0-th order variational parameter $\mathcal{W}^{(0)}$ we obtain the following ansatz  
	\begin{align}
	\psi(\vec{s},t) = \exp \left( \ln \psi_0(\vec{s})  -i E_{\vec{s}} \mathcal{W}^{(0)}  -i  J \sideset{}{'}\sum_{\square}   \frac{\psi_0(\vec{s}_{\square})}{\psi_0(\vec{s})}   \mathcal{W}_{\omega_{\square}(\vec{s})}^{(1)}\right).
	\label{eq:Ansatz_1_order}
	\end{align}
	Such ansatz works for translationally-invariant initial states. For the inhomogeneous initial state we replicate this set of variational parameters for each plaquette or each column depending on the spatial symmetry of the state.
	
	Note that the ansatz (\ref{eq:Ansatz_1_order}) can be explicitly recast to a classical Ising-like spin model with multiple (up to 16) spin interaction terms, $\mathcal{W}^{(1)}_{\omega_{\square}(\vec{s})} = \sum_{\omega=-4}^4 \delta_{\omega, \omega_{\square} (\vec{s})} \mathcal{W}_{\omega}$, where for example 
	$\delta_{\omega=4, \omega_{\square} (\vec{s})} =  P_{1}^- P_{2}^- P_{3}^+ P_{4}^+ P_{a1}^+ P_{a2}^+ P_{a4}^-    P_{b1}^+ P_{b2}^+ P_{b3}^-  P_{c2}^+ P_{c3}^- P_{c4}^-   P_{d1}^+ P_{d3}^- P_{d4}^-  +h.c.$, 
	with $P^{\pm}_i =(1 \pm S^z_i)/2$.
	We emphasize that functions $\omega_{\square}(\vec{s})$ and the corresponding variational parameters $\mathcal{W}_{\omega_{\square}(\vec{s})}^{(1)}$ encode this rather involved spin model with a complex network of local spin connections (i.e. interaction terms) in a much simpler way, while containing exactly the same information. 
	We refer to the ansatz (\ref{eq:Ansatz_1_order}) as to a first-order variational classical network (vCN) for the quantum link model. In the main text we show its simplified version for the case $\psi_0(\vec{s})=const$.
	
	\subsection{Second-order variational ansatz}
	\label{Second-order variational ansatz}
	
	Here we sketch the analogous derivation for the second-order variational ansatz used for the calculations in the present work.
	
	Second-order cumulant expansion is given by
	\begin{align}
	\psi(\vec{s},t) &= \psi_0(\vec{s})  e^{-i E_{\vec{s}} t} \exp \left(  
	-i \int_0^t dt' \frac{\langle \vec{s} | V(t')|\psi_0 \rangle}{\langle \vec{s} | \psi_0 \rangle}  - \right. \nonumber \\
	& \left. - \int_0^t dt' \int_0^{t'} dt'' \left\{ \frac{\langle \vec{s} | V(t') V(t'')|\psi_0 \rangle}{\langle \vec{s} | \psi_0 \rangle} -  
	\frac{\langle \vec{s} | V(t') |\psi_0  \rangle \langle \vec{s} | V(t'') |\psi_0 \rangle}{\langle \vec{s} | \psi_0 \rangle^2}  \right\}
	+ O(V^3)\right)
	\label{eq:Cumulant_expansion_2_order}
	\end{align}
	Using eqs. (\ref{eq:U_t1}), (\ref{eq:U_t2}) and the resolution of identity $\mathds{1} = \sum_{\vec{s}} \ket{\vec{s}} \bra{\vec{s}}$, the first term of the expansion can be recast as
	\begin{align}
	\frac{\langle \vec{s} | V(t') V(t'')|\psi_0 \rangle}{\langle \vec{s} | \psi_0 \rangle}
	=J^2 \sideset{}{''} \sum_{\square', \square''}  \frac{\langle \vec{s}_{\square',\square''} | \psi_0 \rangle}{\langle \vec{s} | \psi_0 \rangle}
	e^{i \lambda \omega_{\square'}(\vec{s})t'}  
	e^{i \lambda \omega_{\square''}(\vec{s}_{\square'})t''}
	\label{eq:2nd-order,1st-term}
	\end{align}
	Here $\sum {}^{''}$ indicates that we sum only over such plaquette pairs that plaquette $\square'$ is flippable in configuration $\vec{s}$ and plaquette $\square''$ is flippable in configuration $\vec{s}_{\square}$; the summation is over the pairs of nearest neighboring and next-nearest neighboring plaquettes (contributions from more distant neighbors are canceled by the second term in the curly brackets in (\ref{eq:Cumulant_expansion_2_order})). After integration over $t'$ and $t''$ eq. (\ref{eq:2nd-order,1st-term}) transforms to
	\begin{align}
	- \int_0^t dt' \int_0^{t'} dt'' \frac{\langle \vec{s} | V(t') V(t'')|\psi_0 \rangle}{\langle \vec{s} | \psi_0 \rangle}
	=J^2 \sideset{}{''} \sum_{\square', \square''} \frac{\langle \vec{s}_{\square',\square''} | \psi_0 \rangle}{\langle \vec{s} | \psi_0 \rangle}  
	f (\omega_{\square'}(\vec{s}), \omega_{\square''}(\vec{s}_{\square'}),t)
	\end{align}
	The pair $(\omega_{\square'}, \omega_{\square''})$ takes (not more than) $9^2=81$ possible values and we promote function $f(\omega_{\square'}, \omega_{\square''},t)$ for each of these values to a variational parameter
	\begin{align}
	f(\omega_{\square'},\omega_{\square''}, t) \equiv \sum_{\omega_1,\omega_2=-4}^4 \delta_{\omega_{\square'},\omega_1} \delta_{\omega_{\square''},\omega_2} f(\omega_1,\omega_2, t) \rightarrow 
	\sum_{\omega_1,\omega_2=-4}^4 \delta_{\omega_{\square'},\omega_1} \delta_{\omega_{\square''},\omega_2} \mathcal{W}^{(2)} _{\omega_1,\omega_2} = \mathcal{W}^{(2)} _{\omega_{\square'(\vec{s})}, \omega_{\square''(\vec{s}_{\square'})}}
	\label{eq:time-to-VP_2-order}
	\end{align}
	Analogously the second term is
	\begin{align}
	-\int_0^t dt' \int_0^{t'} dt'' \frac{\langle \vec{s} | V(t') | \psi_0 \rangle \langle \vec{s} |  V(t'')|\psi_0 \rangle}{\langle \vec{s} | \psi_0 \rangle^2}
	=J^2 \sideset{}{'}\sum_{\square', \square''}  \frac{\langle \vec{s}_{\square'} | \psi_0 \rangle \langle \vec{s}_{\square''} | \psi_0 \rangle}{\langle \vec{s} | \psi_0 \rangle^2} 
	\tilde{f} (\omega_{\square'}(\vec{s}), \omega_{\square''}(\vec{s}),t)
	\end{align}
	Here the prime denotes sum over all pairs of flippable plaquettes. We promote $\tilde{f}(\omega_{\square'}, \omega_{\square''},t)$ to 81 variational parameters $\tilde{\mathcal{W}}^{(2)}$. Finally, the second-order vCN ansatz reads
	\begin{align}
	\psi(\vec{s},t) = \exp \left( \ln \psi_0(\vec{s})  -i E_{\vec{s}} \mathcal{W}^{(0)}  -i  J\sideset{}{'} \sum_{\square}   \frac{\psi_0(\vec{s}_{\square})}{\psi_0(\vec{s})}   \mathcal{W}_{\omega_{\square}(\vec{s})}^{(1)} +\right. \nonumber \\
	\left. +  J^2  \sideset{}{''}\sum_{\square', \square''} \frac{ \psi_0( \vec{s}_{\square',\square''} ) }{ \psi_0(\vec{s})}   \mathcal{W}^{(2)} _{\omega_{\square'(\vec{s})}, \omega_{\square''(\vec{s}_{\square'})}}
	- J^2  \sideset{}{'} \sum_{\square', \square''}  \frac{ \psi_0(\vec{s}_{\square'}) \psi_0(\vec{s}_{\square''})}{ \psi_0(\vec{s})^2}  \tilde{\mathcal{W}}^{(2)} _{\omega_{\square'(\vec{s})}, \omega_{\square''(\vec{s})}}  \right)
	\label{eq:Ansatz_2_order}
	\end{align}

	\newpage 
	
	\section{Time-dependent variational principle}
	\label{Time-dependent variational principle}
	
	The time-dependent variational principle (TDVP) is an approximate Monte Carlo (MC) scheme of evolving the wavefunction provided some suitable ansatz [\onlinecite{Becca-Sorello:book}]. Here we briefly review its main steps.
	
	First, we choose a suitable ansatz $\psi(\vec{s}, \mathcal{W})$ for a given problem (such as the classical network introduced in the previous supplementary section and used in the present work). Here $\mathcal{W}$ denotes a set of complex time-dependent variational parameters $\mathcal{W}(t)=\{\mathcal{W}_0(t), \mathcal{W}_1(t), \dots, \mathcal{W}_K(t)\}$,  where we include an additional ``dummy'' parameter $\mathcal{W}_0$ that fixes the normalization and the global phase of the wavefunction.
	
	Given a state $\ket{\psi_{\mathcal{W}}} = \sum \psi(\vec{s}, \mathcal{W}) \ket{\vec{s}} $ at time $t$ we consider an infinitesimally evolved exact state $|\Phi\rangle= |\psi_{\mathcal{W}}\rangle- i\delta t H|\psi_{\mathcal{W}}\rangle$ and an approximately evolved variational state $|\Psi\rangle= |\psi_{\mathcal{W}}\rangle+\delta t \partial_t|\psi_{\mathcal{W}}\rangle$.
	Let us introduce the total rate of departure of the variational state $\ket{\Psi}$ with respect to the exactly evolved state $\ket{\Phi}$:
	\begin{equation}
	\Delta^2(t) =  \sum_{\{\vec{s}\}} \big| \dot{\Psi}(\vec{s}, t) - \dot{\Phi}(\vec{s},t)\big|^2 = \sum_{\{\vec{s}\}} |\psi(\vec{s},\mathcal{W})|^2\Big| \sum_{k=0}^K O_k(\vec{s})\dot{\mathcal{W}}_k+i E_\mathrm{loc}(\vec{s})\Big|^2,
	\label{eq:error_production_rate}
	\end{equation}
	where $E_\mathrm{loc}(\vec{s})=\langle \vec{s}|H|\psi_{\mathcal{W}}\rangle  / \langle\vec{s}|\psi_{\mathcal{W}} \rangle$ is  the ``local energy'',   $O_k(\vec{s})= \partial \ln \psi(\vec{s}, \mathcal{W}) / \partial \mathcal{W}_k$.
	By minimizing $\Delta^2$ with respect to $\dot{\mathcal{W}}^{\ast}$ we obtain a system of linear first-order differential equations for evolution of the variational parameters $\mathcal{W}(t)$:
	\begin{equation}
	\label{tdvp_eq}
	\sum_{k'=1}^K \mathcal{S}_{k,k'} \dot{\mathcal{W}}_{k'} = -i F_{k},
	\end{equation}
	where $\mathcal{S}$ is the so-called covariance matrix with the matrix elements $\mathcal{S}_{k, k'}=\langle O_k^{\ast} O_{k'}\rangle-\langle O_k^{\ast} \rangle\langle O_{k'}\rangle$, and $F_k=\langle E_\mathrm{loc}O_k^{\ast}\rangle - \langle E_\mathrm{loc}\rangle \langle O_k^{\ast}\rangle $.
	Note that $\mathcal{S}$, $F$ and other observables can be efficiently measured by performing a Monte Carlo sampling of $|\psi(\vec{s},\mathcal{W})|^2$. 
	Solving the system of linear equations (\ref{tdvp_eq}) with respect to $\dot{\mathcal{W}}$ we obtain the time-derivatives $\dot{\mathcal{W}}(t)$, which can be integrated and give us the approximately evolved state $\psi(\vec{s}, \mathcal{W}(t))$. Importantly, the error production rate function (\ref{eq:error_production_rate}) gives us a self-contained method of quantifying the quality of our approximate solution.
	
	\newpage
	
	\section{Benchmarks}
	
	In this Supplementary Section we provide benchmarks for the variational classical network method. The approximate time-evolved state is constructed by using the 2nd order perturbational ansatz (introduced in Suppl. Sec. \ref{Construction of variational classical network Ansatze}), which is evolved using TDVP (Suppl. Sec. \ref{Time-dependent variational principle}). During the evolution we sample various observables using the Monte Carlo method.
	We compare the vCN results with the results of exact diagonalization for the fully flippable sector of $10\times2$ system (the dimension of the Hilbert space is 17906).
	Below we present a) single-plaquette observables, b) correlation functions, and c) error production rate for translationally-invariant initial state $\ket{\rightarrow}_{\mathrm{FF}}$ and a non-uniform initial state $P \ket{\rightarrow}_{\mathrm{FF}}$  defined in the main text.
	
\begin{figure}[h] %
		\centering
		\includegraphics[width=6cm]{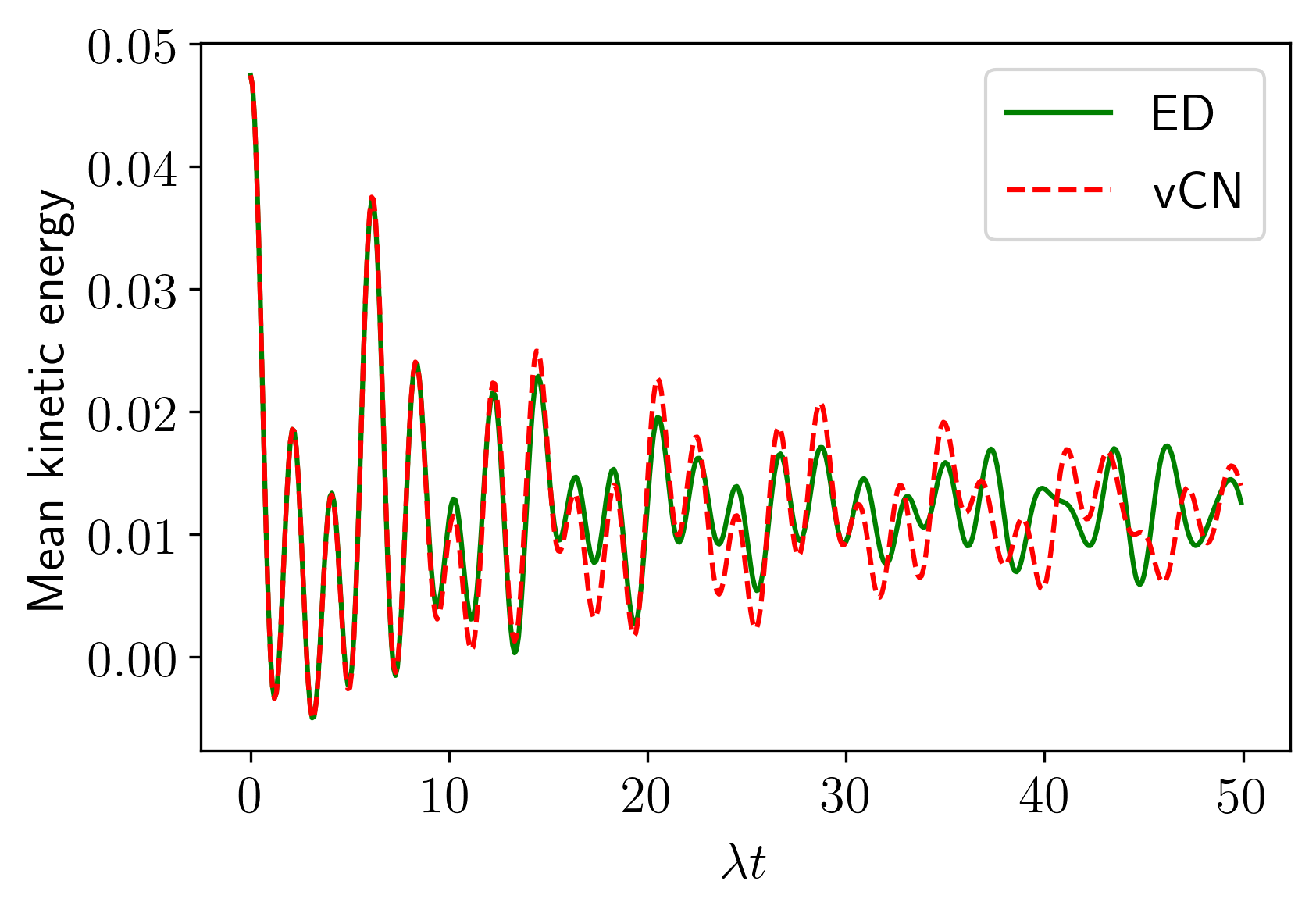}	
		\caption{Mean kinetic energy per plaquette $-J  \langle  U_{\square} + U_{\square}^{\dagger} \rangle$ for the uniform initial state $\ket{\rightarrow}_{\mathrm{FF}}$. Comparison of the exact diagonalization (ED, green) and vCN (dashed red).}
		\label{fig:kinetic_energy}
	\end{figure}

\begin{figure}[tbh] %
		\centering
		\includegraphics[width=0.325\linewidth]{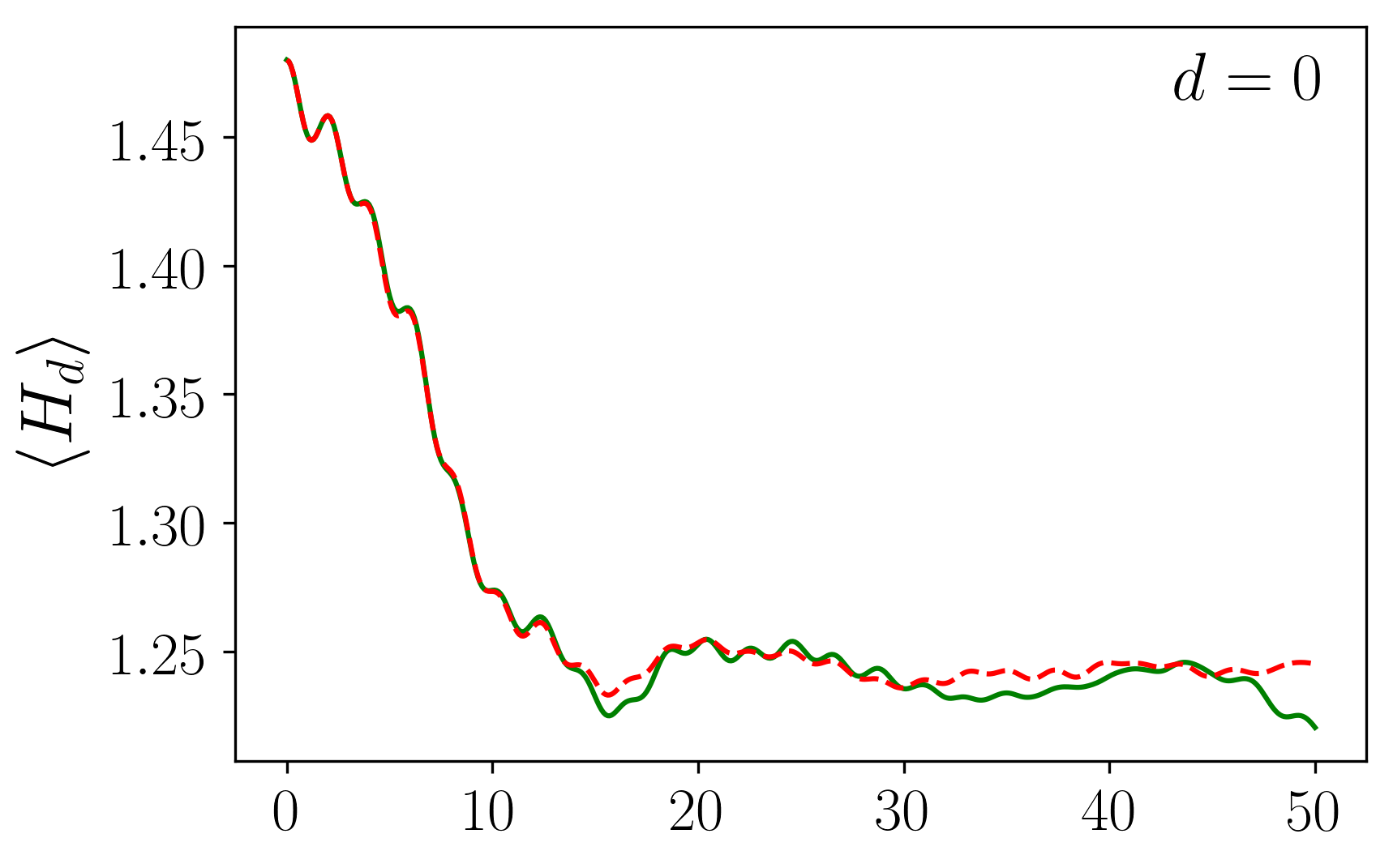}
		\includegraphics[width=0.310\linewidth]{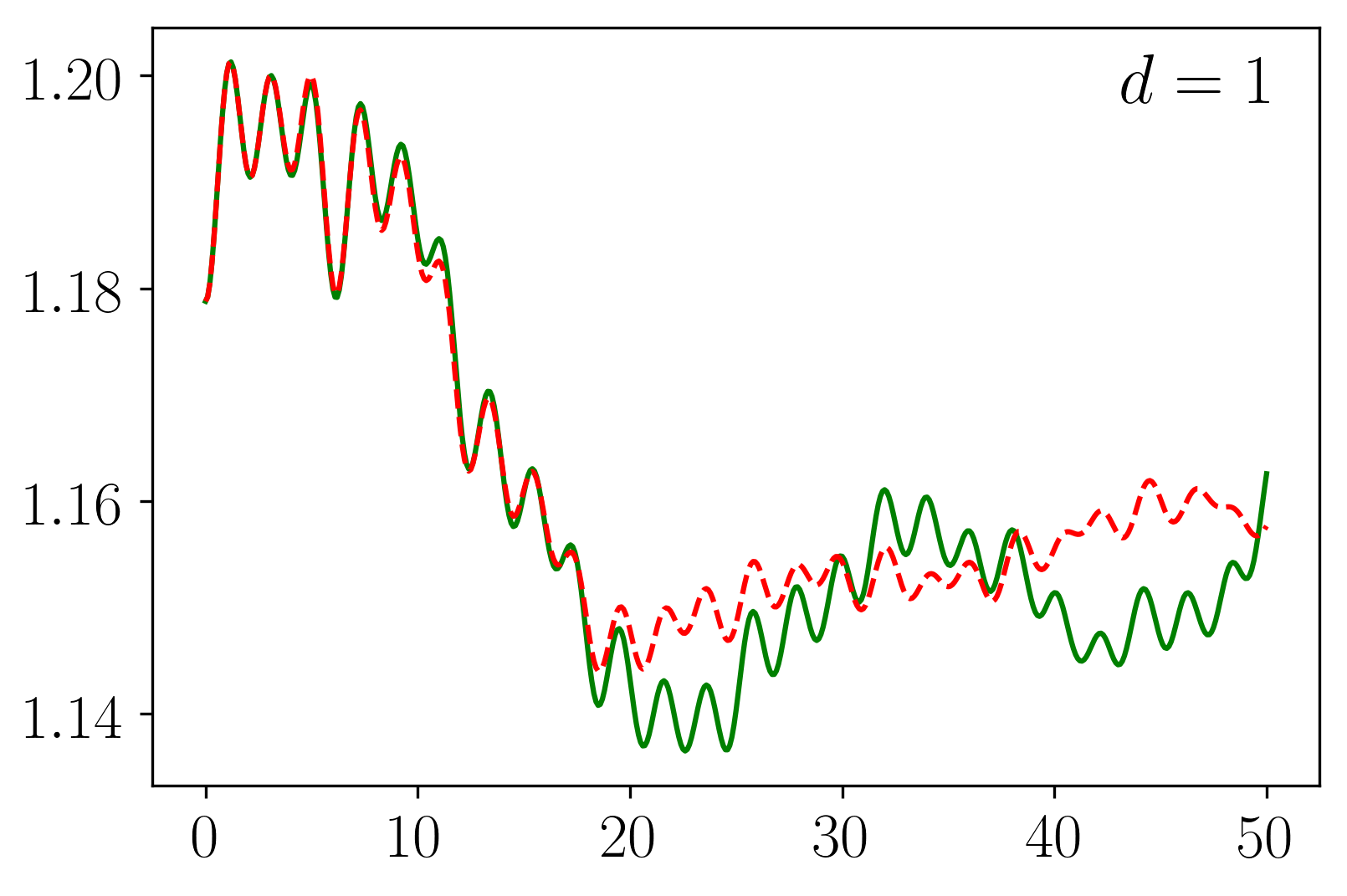}
		\includegraphics[width=0.310\linewidth]{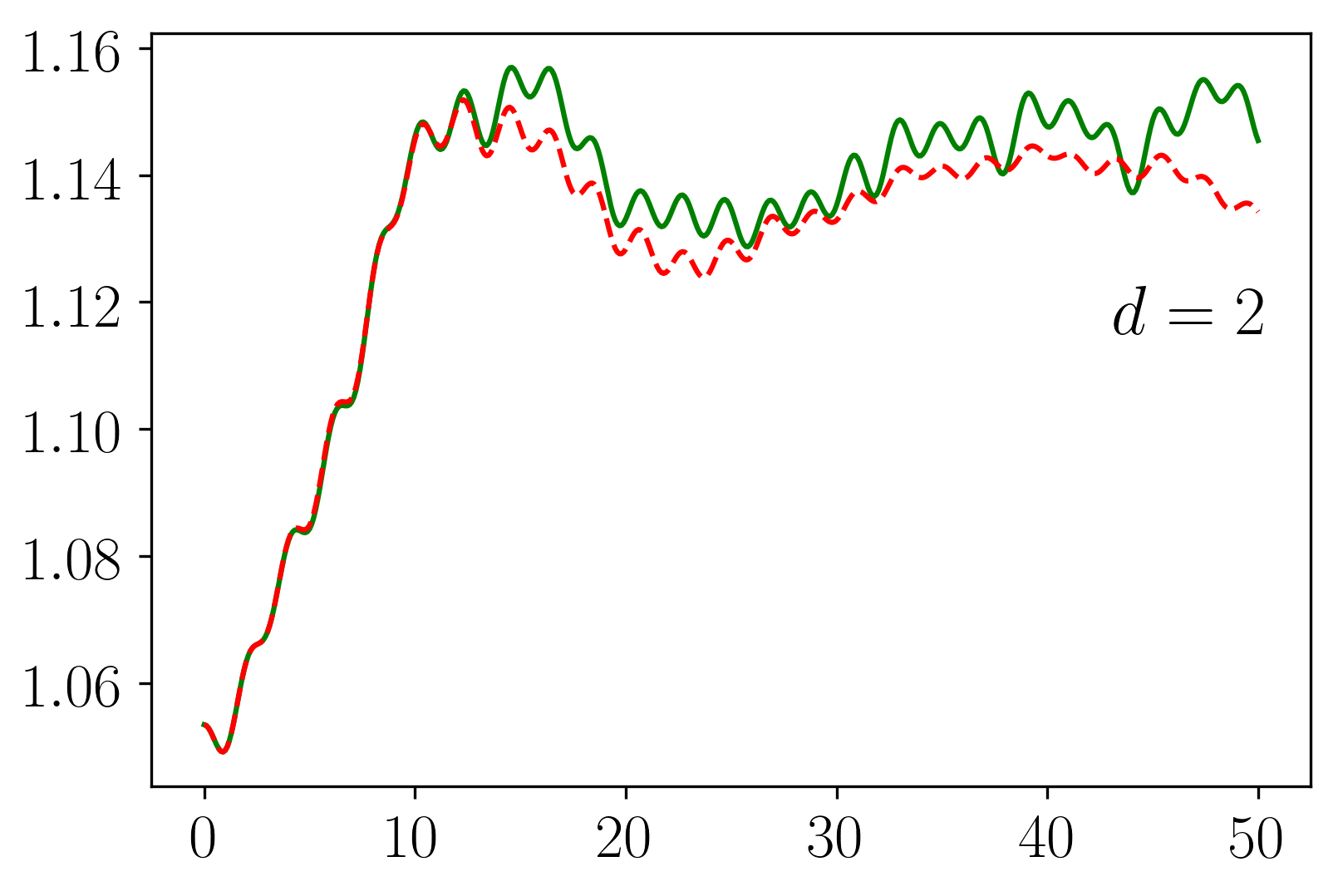}
		\includegraphics[width=0.325\linewidth]{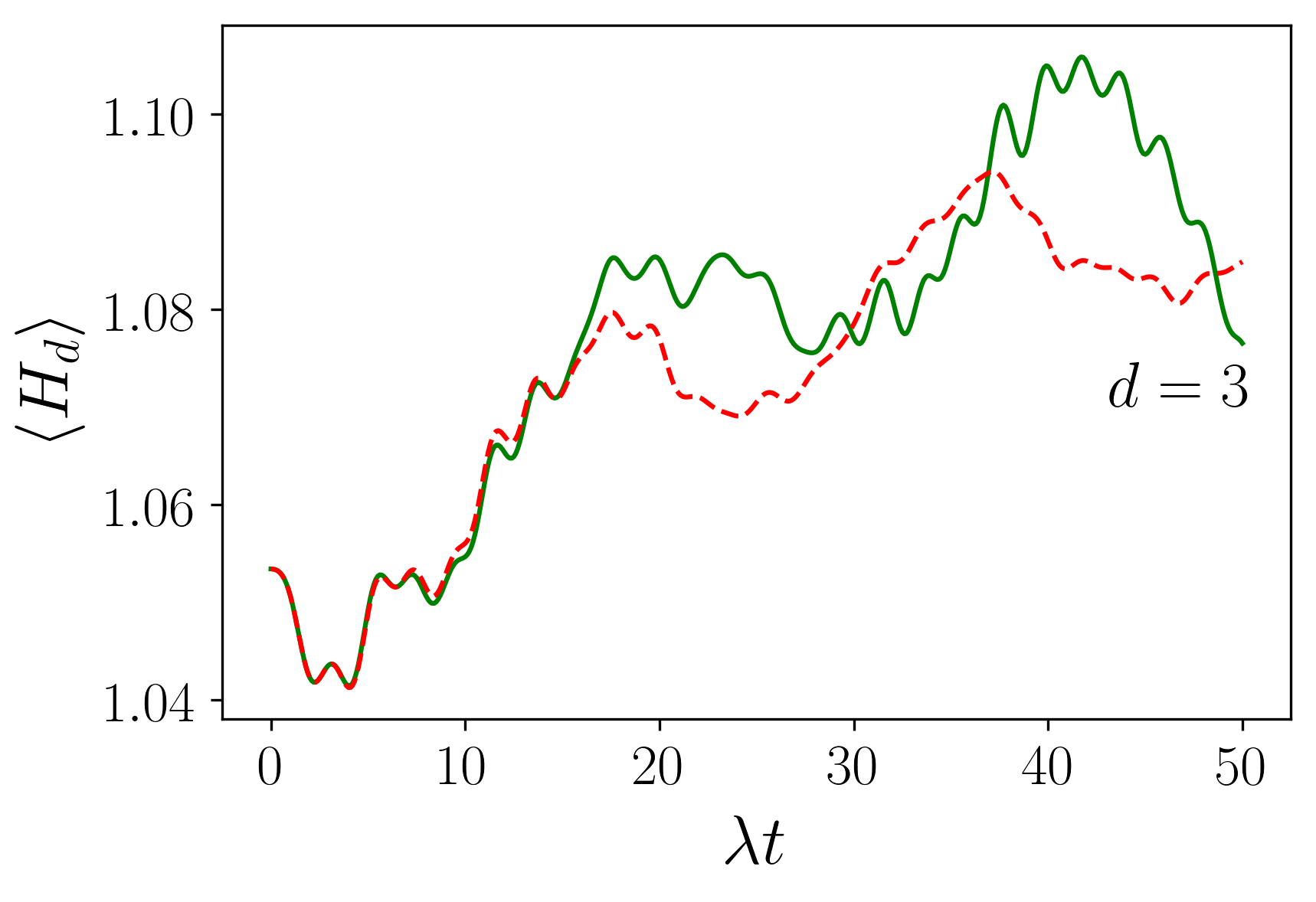}
		\includegraphics[width=0.310\linewidth]{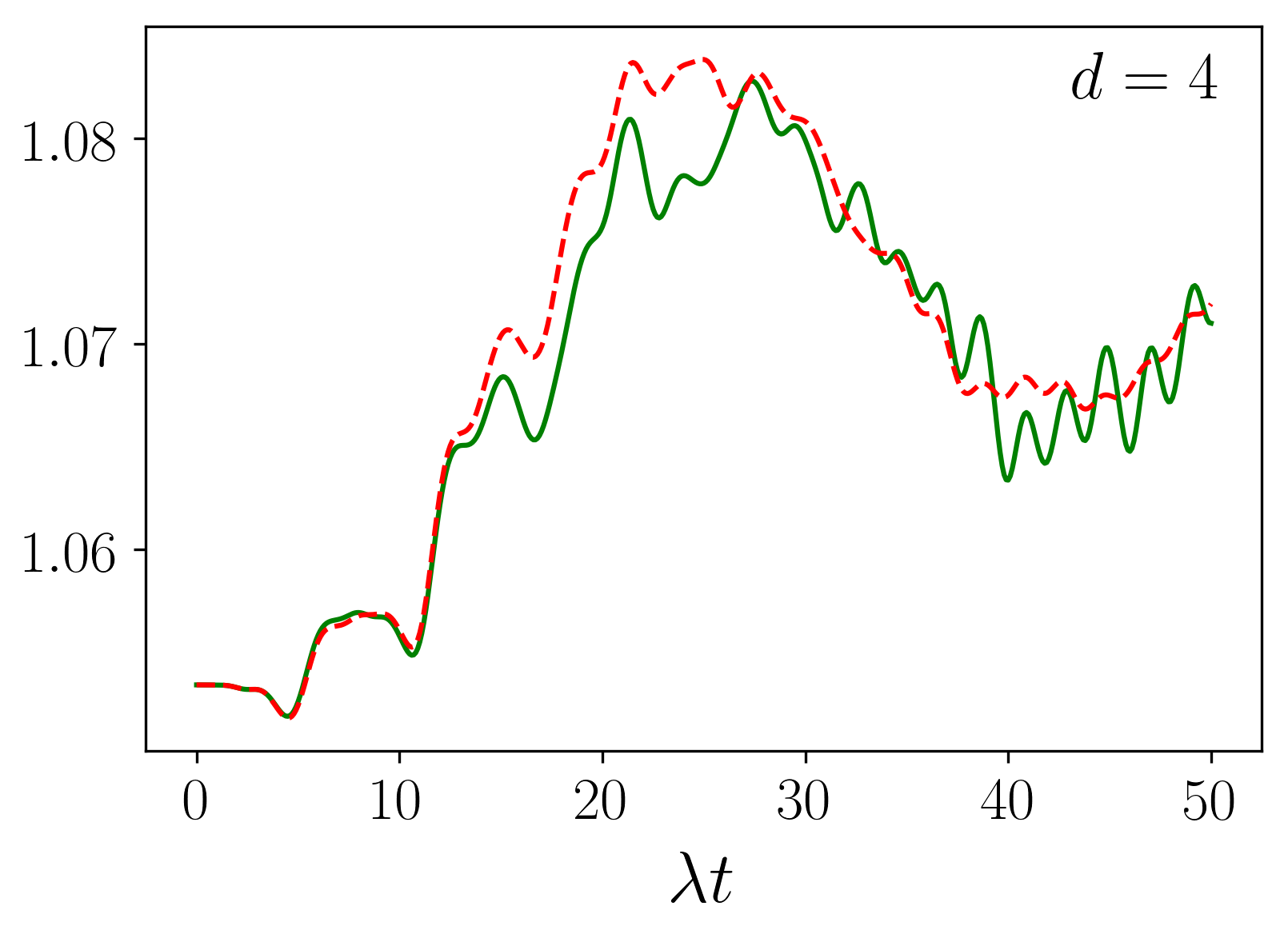} 
		\includegraphics[width=0.310\linewidth]{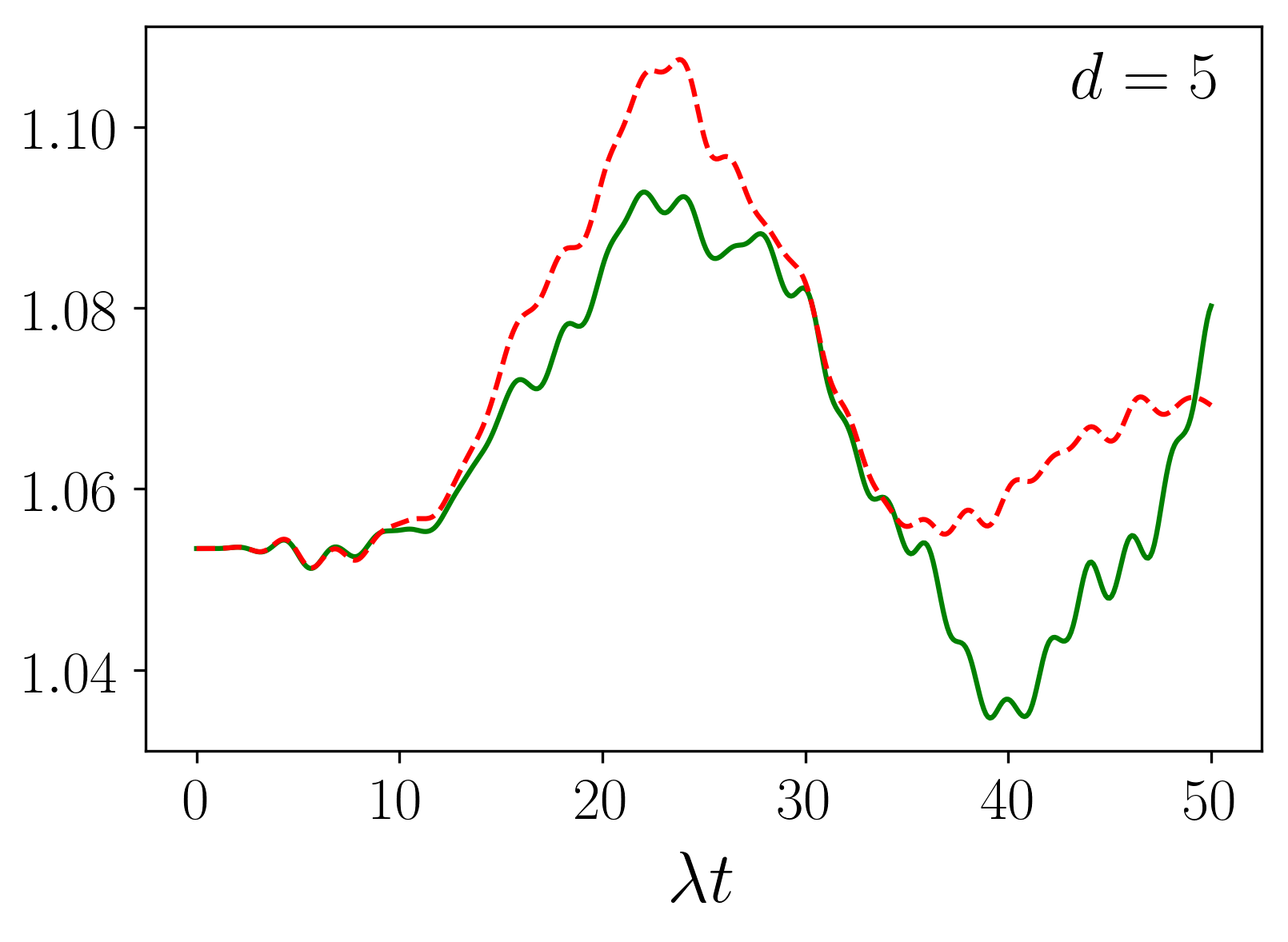}
		\caption{Energy of $d$-th column for $d=0,...,5$ for the non-uniform initial state $P\ket{\rightarrow}_{\mathrm{FF}}$ (defined in the main text) with the excess of energy around $d=0$. Comparison of ED (green) and vCN (dashed red).}
		\label{fig:energy_d-th_column}
	\end{figure}

\begin{figure}[H] %
		\centering
		\includegraphics[width=0.335\linewidth]{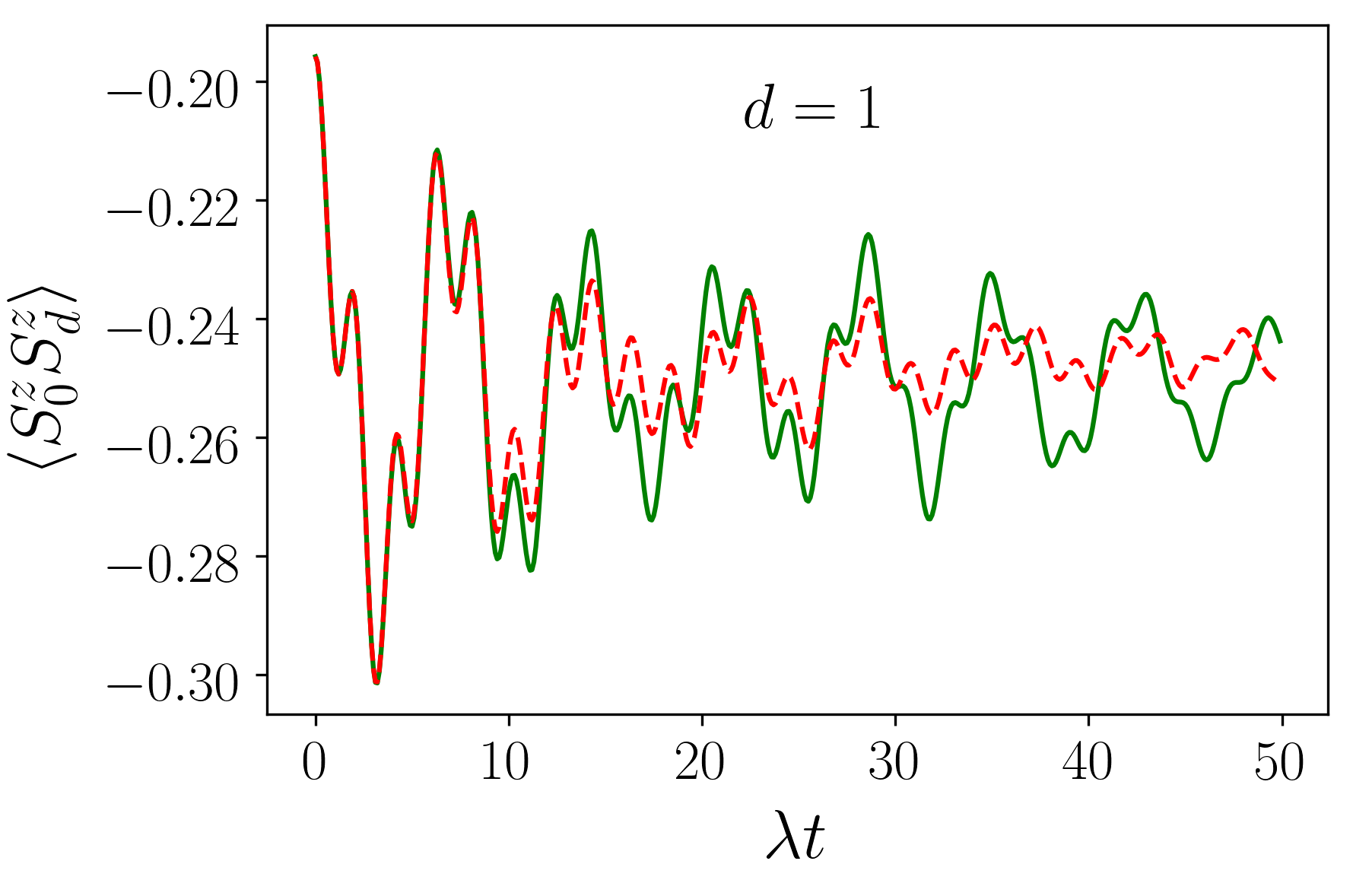}	
		\includegraphics[width=0.32\linewidth]{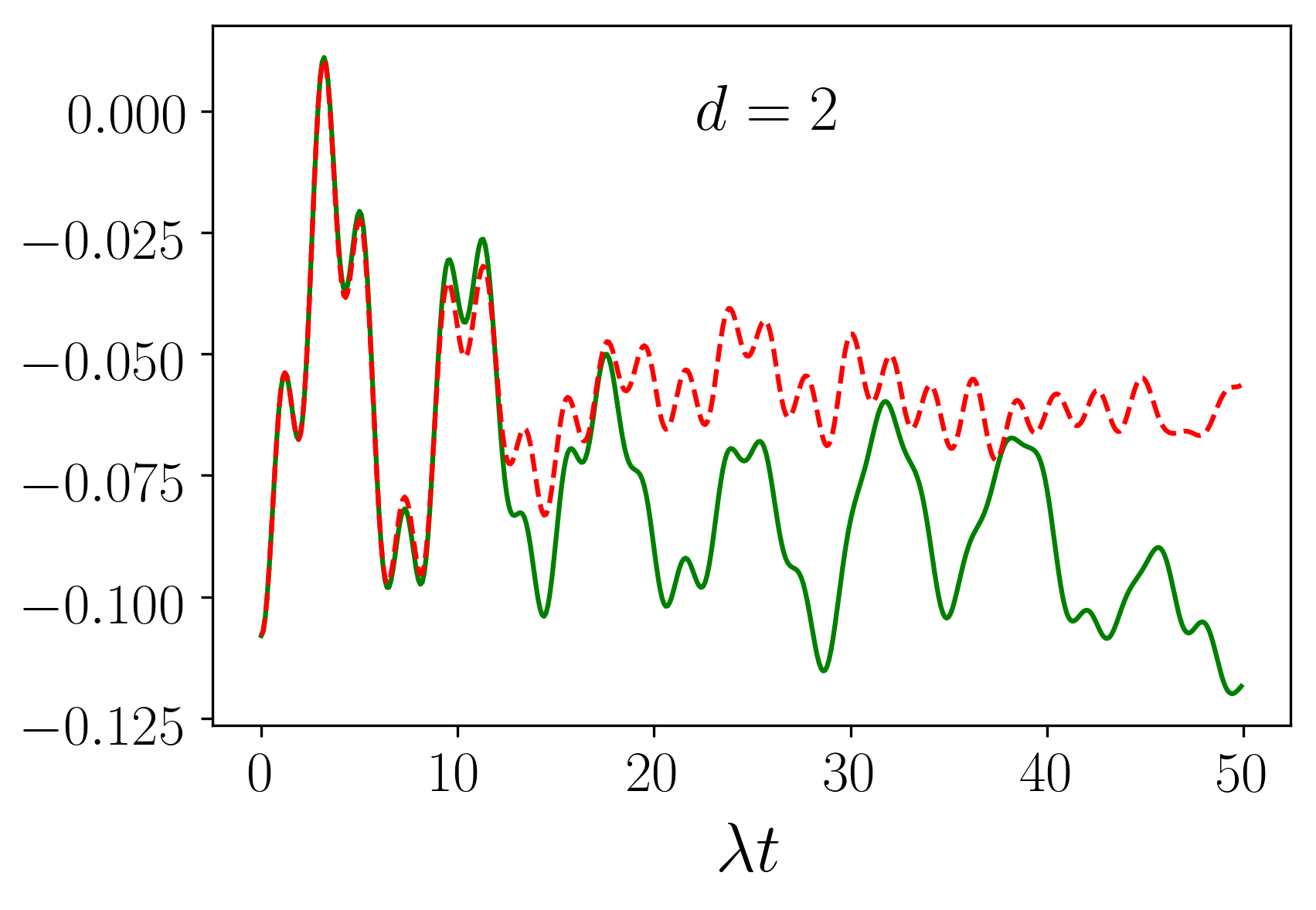}	
		\includegraphics[width=0.32\linewidth]{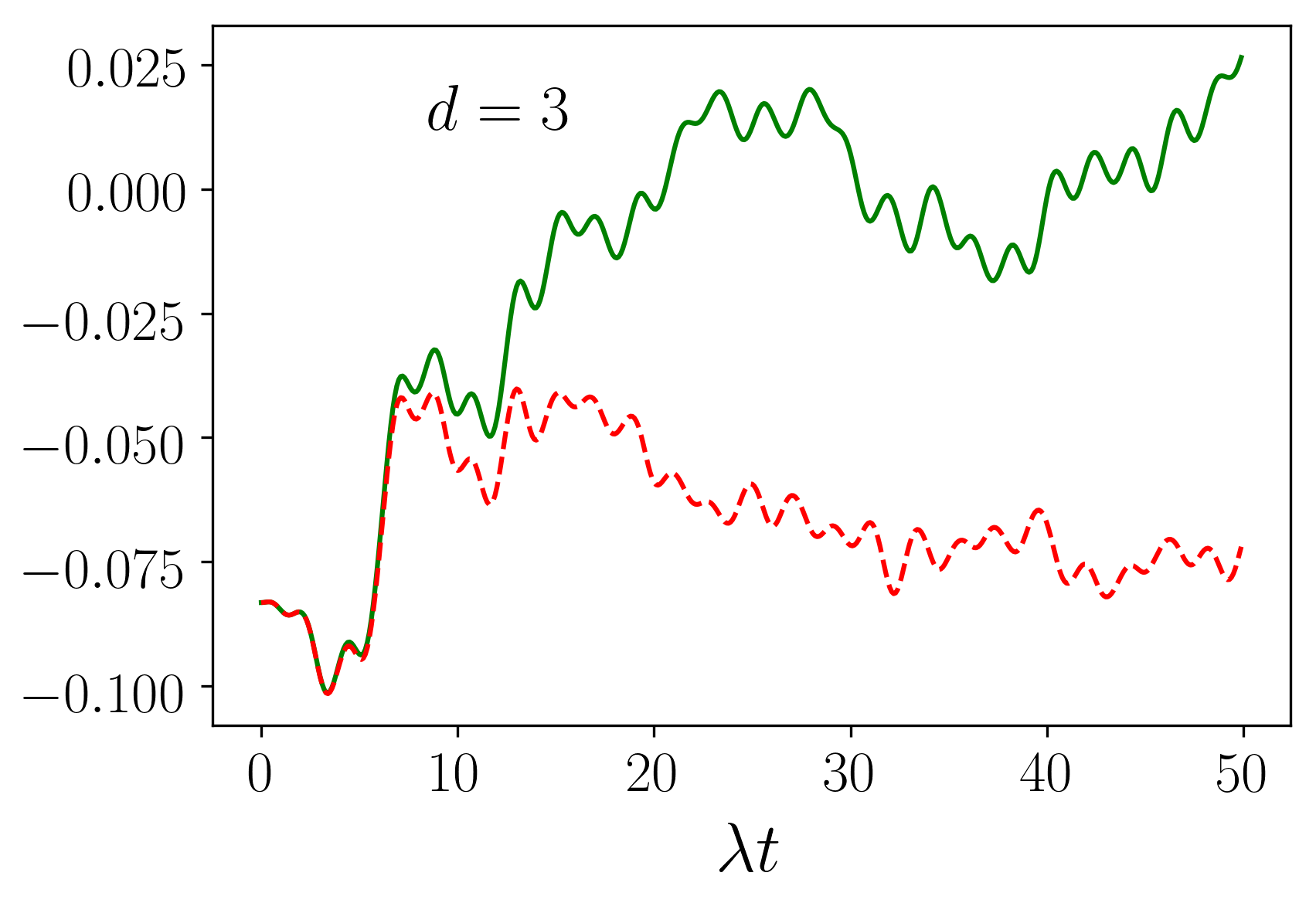}	
		\includegraphics[width=0.34\linewidth]{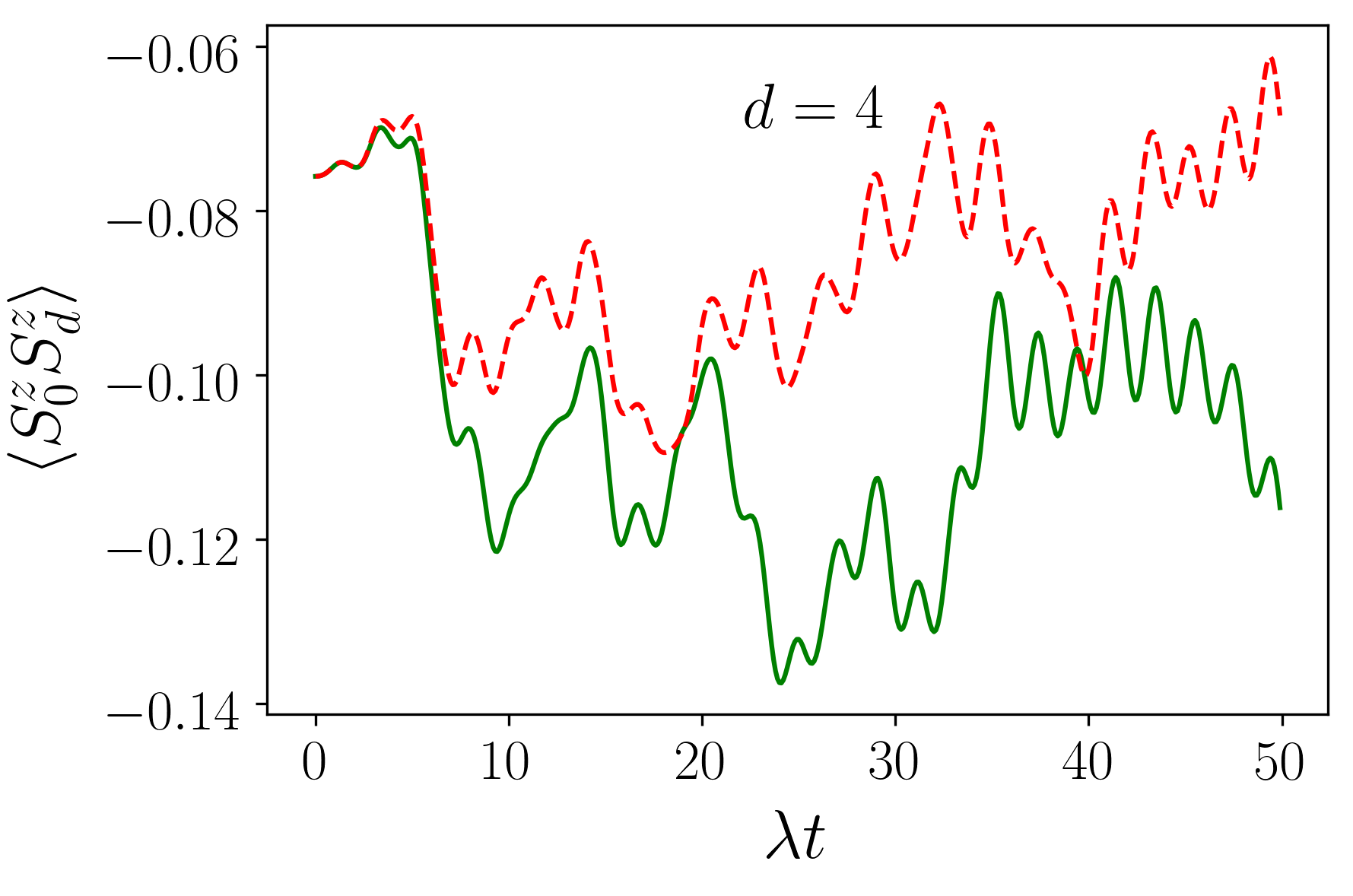}	
		\includegraphics[width=0.32\linewidth]{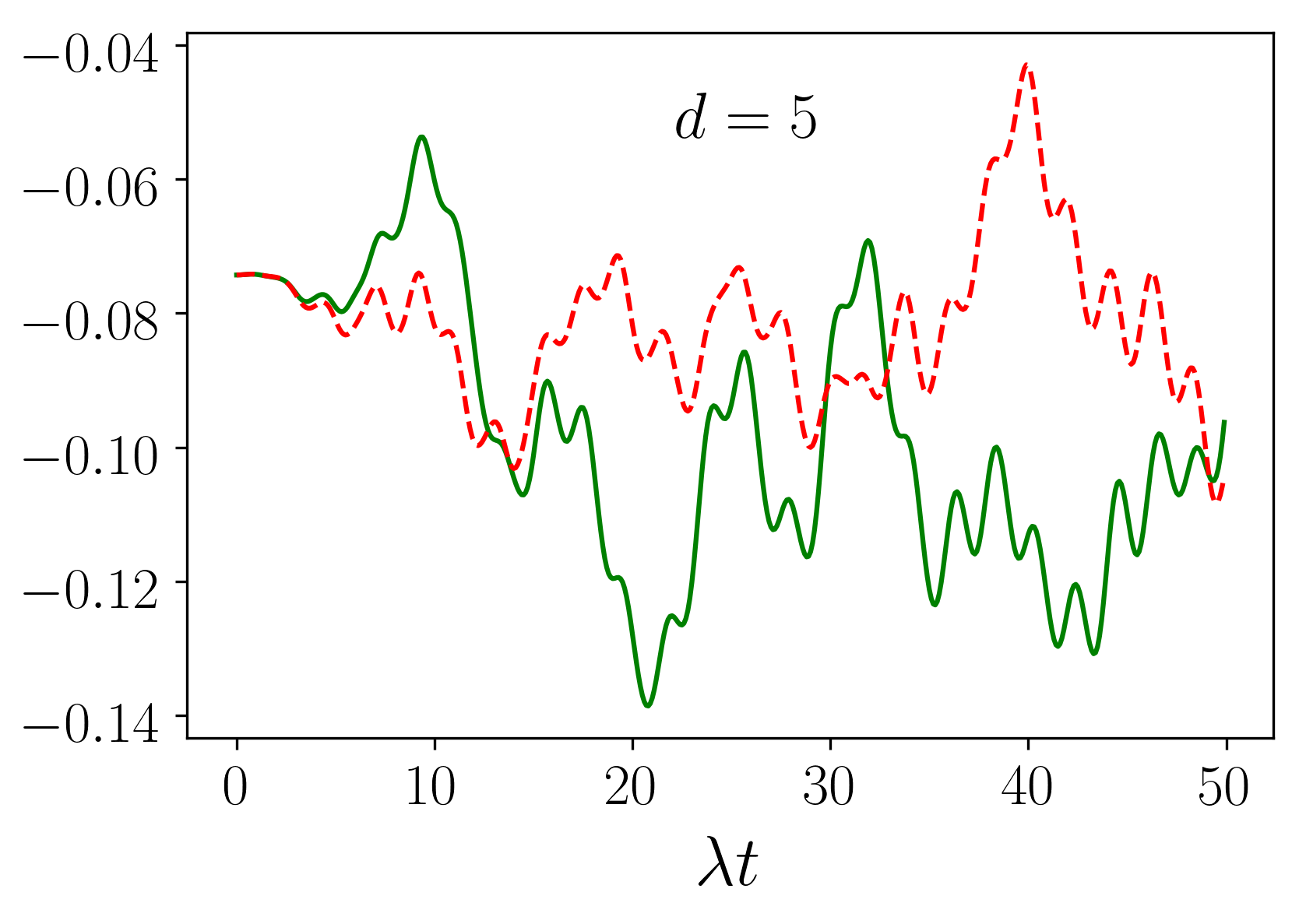}	
		
		\caption{Correlation function $\langle S^z_0 (t) S^z_d (t)\rangle$ for $d=1,...,5$ for the uniform initial state  $\ket{\rightarrow}_{\mathrm{FF}}$.  Comparison of ED (green) and vCN (dashed red).}
		\label{fig:SzSz}
	\end{figure}

\begin{figure}[hbt] %
		\centering
		\includegraphics[width=8cm]{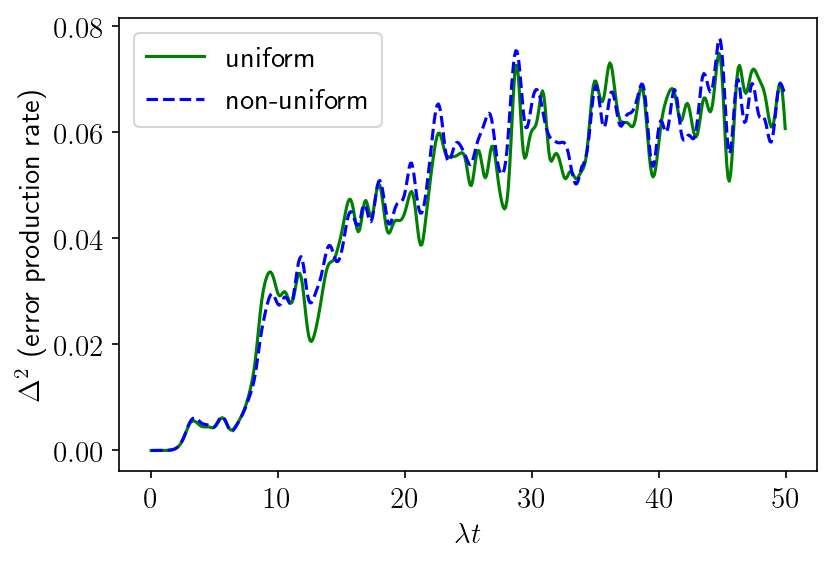}	
		\caption{Error production rate defined by eq. (\ref{eq:error_production_rate}) for uniform state  $\ket{\rightarrow}_{\mathrm{FF}}$  and non-uniform state  $P\ket{\rightarrow}_{\mathrm{FF}}$.}
		\label{fig:error_production_rate}
	\end{figure}

	Figures \ref{fig:kinetic_energy}-\ref{fig:SzSz} show that up to time $\lambda  t\simeq |\lambda/J| \approx 10$, vCN method works very well for both the single-plaquette observables (Figs. \ref{fig:kinetic_energy}-\ref{fig:energy_d-th_column}) and for the long-distance correlators (Fig. \ref{fig:SzSz}), which is also manifested in the fact that the error production rate remains small till  $\lambda t\simeq 10$ and starts significantly grow  after that (Fig. \ref{fig:error_production_rate}). Nevertheless we see that vCN can capture the single plaquette observables up to much longer times, especially the average behavior with smoothed out high-frequency oscillations (note much smaller vertical scale for $d=1-5$ compared to $d=0$ in Fig. \ref{fig:energy_d-th_column}).

	\subsection{Single-plaquette observables}
	\vspace{-0.5cm}
	
	\subsection{Correlation functions}
	
	\subsection{Error production rate}
